\documentclass[twocolumn,showpacs,superscriptaddress,nofootinbib,floatfix]{revtex4}
\usepackage{graphicx}

\begin{document}
\def\Journal#1#2#3#4{{#1} {\bf #2}, #3 (#4)}
\title{\bf The parton bubble model (PBM) is connected to the glasma flux tube model (GFTM), and predicts the ridge and strong CP violation.}
\affiliation{City College of New York, New York City, New York 10031}
\affiliation{Brookhaven National Laboratory, Upton, New York 11973}
\author{S.J.~Lindenbaum}\affiliation{City College of New York, New York City, New York 10031}\affiliation{Brookhaven National Laboratory, Upton, New York 11973}
\author{R.S.~Longacre}\affiliation{Brookhaven National Laboratory, Upton, New York 11973}
\date{\today}
  
\begin{abstract}
In an earlier paper we developed a Parton Bubble Model (PBM) for RHIC, 
high-energy heavy-ion collisions. PBM was based on a substructure of a ring of 
localized bubbles (gluonic hot spots) which initially contain 3-4 partons 
composed of almost entirely gluons. The bubble ring was perpendicular to the
collider beam direction, centered on the beam, at midrapidity, and located on
the expanding fireball surface of Au + Au central collisions (0-10\%) at
$\sqrt{s_{NN}}$=200 GeV. The bubbles emitted correlated particles at kinetic 
freezeout, leading to a lumpy fireball surface. For a selection of charged
particles (0.8 GeV/c $<$ $p_t$ $<$ 4.0 GeV/c), the PBM reasonably 
quantitatively (within a few percent) explained high precision RHIC 
experimental correlation analyses in a manner which was consistent with the 
small observed HBT source size in this transverse momentum range. We 
demonstrated that surface emission from a distributed set of surface sources 
(as in the PBM) was necessary to obtain this consistency. A Glasma Flux Tube 
Model (GFTM) that formed longitudinal flux tubes in the transverse plane of 
two colliding sheets of Color Glass Condensate (CGC), which pass through one 
another had been developed. These sheets create boost invariant flux tubes 
of longitudinal color electric and magnetic fields. A blast wave gives 
the tubes near the surface transverse flow in the same way it gave 
transverse flow to the bubbles in the PBM. In this paper we consider the 
equivalent characteristics of the PBM and GFTM and connect the two models. 
The PBM results discussed above were all obtained without a jet trigger. 
Therefore while investigating the ridge in this paper we add a jet 
trigger to the PBM. When one considers a 3-4 GeV/c transverse momentum 
tagged charged particle in combination with another intermediate 
transverse momentum charged particle in the PBM, the ridge correlation 
is generated and explained. In the GFTM the longitudinal color electric 
and magnetic fields have a non-zero topological charge density 
$F \widetilde F$. These fields cause a local strong CP violation 
which effects charged particle production coming from quarks and 
anti-quarks created in the tube or bubble. For investigating 
these phenomena we use the PBM alone, without a jet trigger. 
We developed four charged particle correlations which show this 
strong CP violation effect and accumulate from bubble to bubble 
independent of whether particles are pushed or pulled 
(by the color electric field), and rotated in a right or left 
handed direction (by the color magnetic field). We also show that 
from previously published analyses of experimental data there 
is already strong evidence for the longitudinal color electric fields,
and predict correlations which can be used to search for color magnetic 
field effects.
\end{abstract}
 
  
\maketitle
 
\section{Introduction and review of models} 

In this paper we discuss the characteristics of the parton bubble model
(PBM)\cite{PBM} and connect them to related characteristics of the glasma
flux tube model (GFTM)\cite{Dumitru}. We will then proceed to use the PBM to
discuss and explain the observed ridge phenomenon by adding appropriate trigger
requirements to the PBM. A successful comparison with experimental data is 
made. We then show how the PBM can be used to search for strong 
CP violations or Chern-Simons topological charge\cite{Simons}. We develop 
four charged particle correlations that can be used to detect the
presence of these color electric and color magnetic fields. Then we show that
existing published analyses can be used to already provide strong experimental
evidence for the longitudinal color electric fields. Our predicted specific
four charged particle correlations can be used to search for experimental
evidence for the color magnetic fields.

\subsection{Parton bubble model development}

Our interest in and the eventual development of the PBM goes back many years
to the early nineteen eighties. Van Hove's work\cite{VanHove} in the early
eighties considered the bubbles as small droplets of quark-gluon plasma (QGP)
assumed to be produced in ultra-relativistic collisions of hadrons and/or
nuclei. His work recognized that experimental detectors only see what is
emitted in the final state at kinetic freezeout. Therefore final state 
predictions of his string model work on plasma bubbles or any other theory
required specific convincing predictions that are observable in the final
state which can be measured by experiment. His work predicted that the
final state rapidity distribution dn/dy of hadrons would exhibit isolated
maxima of width $\Delta y \sim 1$ (single bubble) or rapidity bumps of a
few units (due to those cases producing a few bubbles). These rapidity 
regions would also contain traditional signals of QGP formation. Even this
early work by Van Hove needed his bubbles near the surface of the final
state fireball in order to experimentally measure the particles emitted 
in the final state at kinetic freezeout. We and other searched for these 
Van Hove bubbles, but no one ever found any significant evidence for them.

Our next paper\cite{2000} was our final attempt to make a theoretical
treatment of the single bubble case (similar to the Van Hove case) which
could be experimentally verified. It is possible that with enough
statistics one in principle could find single to a few bubble events.
We developed a number of event generators which could possibly provide
evidence for striking signals resulting from these bubbles. One should note
that in that paper we had included in Sec. 11 mathematical expressions
describing how charged pions are effected by strong color electric and
color magnetic fields which are present and parallel in a CP-odd bubble
of metastable vacuum. This section was motivated by the work of
Kharzeev and Pisarski\cite{Pisarski}. For example eq. 3-4 of the section
show the boosts of the $\pi^+$ and the $\pi^-$ we estimated. Thus in
2000 we were already interested in and were investigating and publishing
predictions for this phenomenon.

When we considered the RHIC quantum interference data in 2000\cite{Adler} it 
became clear that the fireball surface was rapidly moving outward. This implied
a very large phase space region covered by the fireball, and also implied that 
it would be very unlikely for there to be only one isolated bubble 
(small sourcesize) with a  large amount of energy sitting on the surface. 
It seemed more likely that there would be many bubbles or gluonic hot spots 
around the expanding surface. This led to a paper\cite{themodel} which was 
our earlier version of Ref.\cite{PBM}. In that paper we concluded that the 
behavior of the Hanbury-Brown and Twiss (HBT) measurements\cite{Adler,HBT} 
should be interpreted as evidence of a substructure of bubbles 
located on the surface of the final state fireball in the central 
rapidity region at kinetic freezeout. The HBT radii were decreasing 
almost linearly with transverse momentum and implying to us a source 
size of $\sim$2 fm radii bubbles were on the surface and could be 
selected for if we considered transverse momenta above 0.8 GeV/c. These
momenta would allow sufficient resolution to resolve individual bubbles
of $\sim$2 fm radii. We further concluded that these HBT quantum interference
observations were likely due to phase space focusing of the bubbles 
pushed by the expanding fireball. Thus the HBT measurements of source
size extrapolated to transverse momenta above 0.8 GeV/c were images of 
the bubbles. The HBT correlation has the property of focusing 
these images on top of each other for a ring of bubbles transverse to 
and centered on the beam forming an average HBT radius. 

Thus our model was a ring of bubbles sitting on the freezeout surface with 
average size $\sim$2 fm radius perpendicular to the beam at mid-rapidity. The 
phase space focusing would also lead to angular correlations between 
particles emitted by the bubbles and be observable. Fig. 1 of that paper shows 
such a bubble geometry. For the background particles that account for 
particles in addition to the bubble particles we used unquenched 
HIJING\cite{hijing} with its jets removed. We assumed that most of the jets 
were eliminated from the central events because of the strong jet quenching 
observed at RHIC\cite{quench1}. We investigated the feasibility of using 
charged particle pair correlations as a function of angles which should 
be observable due to the phase space focusing of the particles coming from 
individual bubbles. At that point we fully expected that these correlations 
would be observable in the STAR detector\cite{star} at RHIC if one analyzed
central Au + Au collisions. Correlation analyses are powerful tools in 
detecting substructures. Historically substructures have played an 
important role in advancing our understanding of strong (non perturbative) 
interactions. Our earlier paper explained the general characteristics 
of the angular correlation data, and was also consistent with HBT 
measurements in a qualitative manner. This motivated us to develop 
a reasonably quantitative model, the parton bubble model\cite{PBM} 
which is discussed in the following.

\subsection{Parton bubble model\cite{PBM}} 

    In this publication\cite{PBM} we developed a QCD inspired parton 
bubble model (PBM) for central (impact parameter near zero) high energy heavy 
ion collisions at RHIC. The PBM is based on a substructure consisting of a 
single ring of a dozen adjoining 2-fm-radius bubbles (gluonic hot spots)
transverse to the collider beam direction, centered on the beam, and located 
at or near mid-rapidity. The ring resides on the fireball blast wave surface
(see Fig. 1 of Ref.\cite{PBM}). We assumed these bubbles are likely the 
final state result of quark-gluon-plasma (QGP) formation since the energy
densities produced experimentally are greater than those estimated as
necessary for formation of a quark-gluon-plasma. Thus this is the geometry 
for the final state kinetic freezeout of the QGP bubbles on the surface of 
the expanding fireball treated in a blast wave model. 

The twelve bubble ring creates the average behavior of bubble formation 
driven by the energy density near the surface of the expanding fire ball of 
the blast wave; that forms the final state surface bubbles that emit the final
state particles at kinetic freezeout. One should note that the blast wave
surface is moving at its maximum velocity at freezeout (3c/4). For central 
events each of the twelve bubbles have 3-4 partons per bubble each at a fixed 
$\phi$ for a given bubble. The transverse momentum ($p_t$) distribution of the 
charged particles is similar to pQCD but has a suppression at high $p_t$ like 
the data.

The bubble ring radius of our model was estimated by blast wave, HBT  and other
general considerations to be approximately 8 fm. The bubbles emit correlated 
charged particles at final-state kinetic freezeout where we select a $p_t$ 
range (0.8 GeV/c $<$ $p_t$ $<$ 4.0 GeV/c) in order to increase signal to 
background. The 0.8 GeV/c $p_t$ cut increases the resolution to allow resolving
individual bubbles which have a radius of $\sim$2 fm. This space momentum 
correlation of the blast wave provides us with a strong angular correlation 
signal. PYTHIA fragmentation functions\cite{pythia} were used for the bubbles 
fragmentation that generate the final state particles emitted from the bubbles.
The PBM explained the high precision Au + Au central (0-10\%) collisions at
$\sqrt{s_{NN}} =$ 200 GeV\cite{centralproduction} (the highest RHIC energy).
The PBM fit to the angular correlation data was reasonably quantitative to
within a few percent (see Sec. 4 of Ref.\cite{PBM}).

The correlation functions we employed (like the HBT correlation functions) have
the property that for the difference in angles (difference of momentum for HBT)
these correlations will image all 12 bubbles on top of each other. This leads
to average observed angles of approximately $30^\circ$ in $\Delta \phi$, 
$70^\circ$ in $\Delta \eta$, originating from a source size of about $\sim$2 fm
radius which is consistent with the HBT correlation. Thus the PBM generates
$\Delta \phi$ vs $\Delta \eta$ charged-particle-pair correlations for charged
particles with $p_t$ in the range 0.8 Gev/c to 4.0 GeV/c as displayed in 
Fig. 1. See Sec. 4 of Ref.\cite{PBM} for a reasonably quantitative
successful comparison with data. Furthermore the model results were consistent 
with the Hanbury-Brown and Twiss (HBT) observations\cite{HBT} that the
observed source radii determined by quantum statistical interference were 
reducing by a considerable factor with increasing transverse momentum ($p_t$). 
The HBT radii are interpreted to reduce from $\sim$6 fm at $p_t$ 
$\sim$0.2 GeV/c to $\sim$2 fm at $p_t$ $\sim$1 GeV/c for our $p_t$ range. The 
generally accepted explanation for this behavior is that as $p_t$ increases 
radial flow increasingly focused the viewed region of the final state into 
smaller volumes. If just one small HBT size bubble were emitting all the 
correlated particles, this phenomena would lead to large spikes of particles 
emitted at one limited $\phi$ angular region in individual events. This is 
definitely not observed in the Au + Au or other collision data at RHIC. 
Therefore, a distributed ring of small sources around the beam as assumed in 
the PBM is necessary to explain both the HBT results\cite{HBT} and the 
correlation data\cite{centralproduction}. The particles emitted from the
same bubble are virtually uncorrelated to particles emitted from any other
bubble except for momentum conservation requirements. An away side peak
in the total correlation is built up from momentum conservation between 
the bubbles.

The PBM was also recently extended to PBME\cite{PBME} which is identical to the
PBM for central collisions (0-5\%). For centralities running from 30-80\% jet
quenching is not strong enough to make jets negligible therefore,
a jet component was added which was based on HIJING calculations. This jet 
component accounts for more of the correlation as one moves toward
peripheral bins, and explains all of it for the most peripheral collisions. 
The PBME explained in a reasonably quantitative manner (within a few percent) 
the behavior of the recent quantitative experimental analysis of 
charge pair correlations as a function of 
centrality\cite{centralitydependence}. This further strengthened 
the substantial evidence for bubble substructure. The agreement of the PBM and
the PBME surface emission models with experimental analyses strongly implied
that at kinetic freezeout the fireball was dense and opaque in the central 
region and most centralities (except the peripheral region) in the intermediate
transverse momentum region (0.8 $<$ $p_t$ $<$ 4.0 GeV/c). Thus we conclude
that the observed correlated particles are both formed and emitted from or 
near the surface of the fireball. In the peripheral region the path to the 
surface is always small.

In the PBM and the PBME the bubbles are produced in large numbers, are most 
symmetric, and have unusual features in the most central collisions that 
can be related to the GFTM. Therefore the central region PBM is the model 
most suitable for comparing to the GFTM. 

The PBM fit to the central production Au + Au correlation 
data\cite{centralproduction} which was observed in the STAR TPC detector
reasonably quantitatively fit the data. The average observed angles were 
approximately $30^\circ$ in $\Delta \phi$ and $70^\circ$ in $\Delta \eta$, 
originating from a source size of $\sim$2 fm radius. Thus the PBM generates  
$\Delta \phi$ vs $\Delta \eta$ charged-particle-pair correlations for the 
charged particles with $p_t$ in the range 0.8 GeV/c to 4.0 GeV/c as displayed 
in Fig. 1 of the present paper. See Sec. 4 of Ref.\cite{PBM} for a 
reasonably quantitative successful comparison with data 
(within a few percent of the observed correlations).

\begin{figure*}[ht] \centerline{\includegraphics[width=0.800\textwidth]
{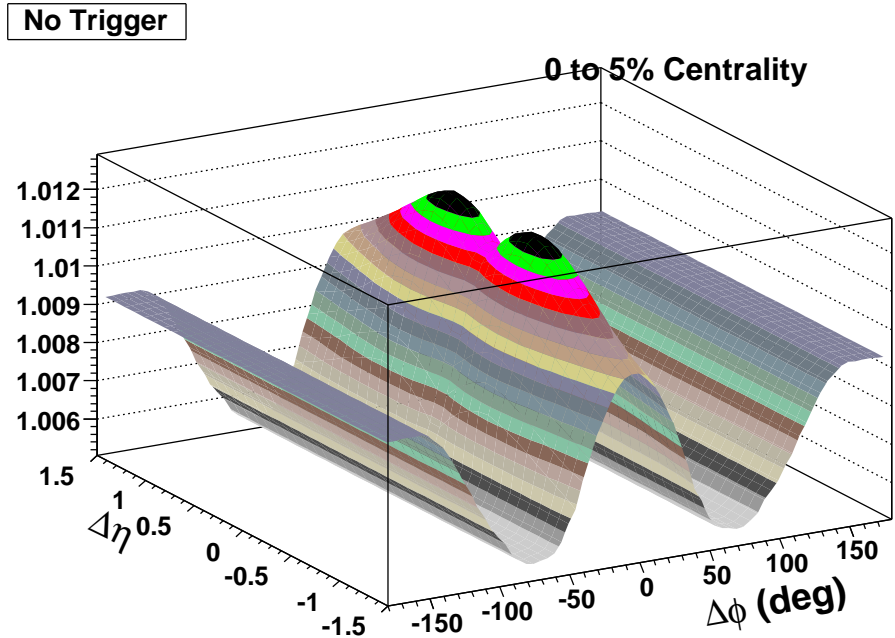}} \caption{The CI (sum of all charged-particle-pairs) correlation 
for the 0-5\% centrality bin with a charged particle $p_t$ selection 
(0.8 $<$ $p_t$ $<$ 4.0 GeV/c), generated by the PBM. It is  plotted as a two 
dimensional $\Delta \phi$ vs. $\Delta \eta$ perspective plot.}
\label{fig1}
\end{figure*}
                                                                              
The CI correlation used in Fig. 1 and the PBM\cite{PBM} is defined below.
The two different types of charged-particle-pair correlations are the 
unlike-sign charge pairs (US) and the like-sign charge pairs (LS). The
charge-independent correlation (CI) is defined as the correlation made up of
charged-particle-pairs independent of what the sign is, which is the average
of the US plus the LS correlations. Thus CI = (US + LS)/2. Therefore the
CI is the total charge pair correlation observed in the experimental
detection system within its acceptance.

We utilize a two particle correlation function in the two dimensional (2-D)
space\footnote{$\Delta \phi = \phi_1 - \phi_2$ where $\phi$ is the azimuthal
angle of a particle measured in a clockwise direction about the beam.
$\Delta \eta = \eta_1 - \eta_2$ which is the difference of the psuedorapidity
of the pair of particles} of $\Delta \phi$ versus $\Delta \eta$ (see 
Ref.\cite{PBM} Sec. 2.1). The PBM generated the two dimensional 
differences for a pair of particles $\Delta \phi$ versus $\Delta \eta$ 
contributing to the correlation function. The sum over all pairs coming from 
the same event and then summed over all events is our numerator in the 
correlation function. We then divide this numerator by a sum over all pairs 
coming from created mixed pairs where, the two particles are produced 
in different events, and then summed over all mixed events. The sum over all 
the particle pair entries in the mixed event denominator is normalized to 
equal the sum over all particle pairs in the numerator. This rescales the CI 
such that the ratio of the numerator divided by the denominator has a mean 
of 1.

The CI was generated for $\sqrt{s_{NN}}$ = 200 GeV central (0-5\%) Au + Au
collisions in the $p_t$ range 0.8 GeV/c to 4.0 GeV/c at RHIC. When
the CI was compared to the corresponding experimental 
analyses\cite{centralproduction,centralitydependence} a reasonably quantitative
agreement within a few percent of the observed CI was attained. See 
Ref.\cite{PBM} Sec. 4.2, Ref.\cite{PBME} Sec. V, and 
Ref.\cite{centralitydependence} Sec. VI B\footnote{The STAR collaboration 
preferred to use the conventional definition of CI = US + LS which is larger by
a factor of 2 than the CI definition used in this present paper that is
directly physically meaningful.}. The CI correlation for the 0-5\% centrality 
bin is shown in Fig. 1. This analysis and our previous work\cite{PBM,PBME} 
was done without using a jet trigger. In fact jet production in 
the central region was negligible due to strong jet 
quenching\cite{quench1,quench2,quench3}. As one can see in Fig. 1 the 
central CI correlation of $\Delta \phi$ on the near side 
($\Delta \phi$ $<$ $90^\circ$), is sharp and approximately jetlike 
at all $\Delta \eta$. In contrast the $\Delta \eta$ dependence 
is relatively flat. In the central production the average observed 
angles are approximately $30^\circ$ in $\Delta \phi$ and $70^\circ$ 
in $\Delta \eta$. It is of interest to note that the sharp jetlike 
collimation in $\Delta \phi$ persists at all centralities (0-80\%). 
The elongation in $\Delta \eta$ persists in the 0-30\% centrality range 
and then decreases with decreasing centrality becoming jetlike in the 
peripheral bins. These characteristics are in reasonable quantitative agreement
with the fits of the PBM and the PBME to experimental data 
analyses\cite{PBM,centralproduction,PBME,centralitydependence}.

The difference of the US and LS correlations is defined as the charge-dependent
(CD) correlation (CD = US - LS). The 2-D experimental CD correlation for 
centralities (0-80\%) has a jetlike shape which is consistent with PYTHIA jets 
(vacuum fragmentation) for all centralities. This clearly implies
that both particle hadroniztion and emission occur from the fireball
surface region as explained in Ref.\cite{PBME} Sec. II.

\subsection{Glasma flux tube model} 

A glasma flux tube model (GFTM)\cite{Dumitru} that had been developed considers
that the wavefunctions of the incoming projectiles, form sheets of color glass 
condensates (CGC)\cite{CGC} that at high energies collide, interact, and 
evolve into high intensity color electric and magnetic fields. This collection 
of primordial fields is the Glasma\cite{Lappi,Gelis}, and initially it is 
composed of only rapidity independent longitudinal color electric and magnetic 
fields. An essential feature of the Glasma is that the fields are localized 
in the transverse space of the collision zone with a size of 1/$Q_s$. $Q_s$ is
the saturation momentum of partons in the nuclear wavefunction. These
longitudinal color electric and magnetic fields generate topological 
Chern-Simons charge\cite{Simons} which becomes a source for particle 
production.

The transverse space is filled with flux tubes of large longitudinal extent 
but small transverse size $\sim$$Q^{-1}_s$. Particle production from a flux 
tube is a Poisson process, since the flux tube is a coherent state. As the
partons emitted from these flux tubes locally equilibrate, transverse flow
builds due to the radial flow of the blast wave\cite{Gavin}. The flux tubes
that are near the surface of the fireball get the largest radial flow and
are emitted from the surface. As in the parton bubble model these partons
shower and the higher $p_t$ particles escape the surface and do not interact.
These flux tubes are in the present paper considered strongly connected to the 
bubbles of the PBM. The method used to connect the PBM and the GFTM is
described and discussed in the next subsection ID.
 
$Q_s$ is around 1 GeV/c thus the transverse size of the flux tube is about 
1/4 fm. The flux tubes near the surface are initially at a radius $\sim$5 fm. 
The $\phi$ angle wedge of the flux tube is $\sim$1/20 radians or 
$\sim$$3^\circ$. Thus the flux tube initially has a narrow range in $\phi$. The
large width in the $\Delta \eta$ correlation which in the PBM depended on the 
large spread in $\Delta \eta$ of the bubble partons results from the 
independent longitudinal color electric and magnetic fields that created the 
Glasma flux tubes. How much of these longitudinal color electric and magnetic 
fields are still present in the surface flux tubes when they have been pushed 
by the blast wave will be a speculation of this paper for measuring strong CP
violation? 

It has been noted that significant features of the PBM that generated final 
state correlations which fit the experimentally observed correlation 
data\cite{PBM,centralproduction,PBME,centralitydependence} are similar to 
those predicted by the GFTM\cite{Dumitru}. Therefore in the immediately
following subsection we assume a direct connection of the PBM and the GTM, 
give reasons to justify it, and then discuss its consequences, predictions
and successes.

\subsection{The connection of the PBM and the GFTM}

The successes of the PBM have strongly implied that the final state surface
region bubbles of the PBM represent a significant substructure. In this
subsection we show that the characteristics of our PBM originally
developed to fit the precision STAR Au + Au correlation data in a manner 
consistent with the HBT data; implies that the bubble substructure we 
originally used to fit these previous data is closely related to the GFTM.

A natural way to connect the PBM bubbles to the GFTM flux tubes is to 
assume that the final state at kinetic freezeout of a flux tube is a PBM
final state bubble. Thus the initial transverse size of a flux tube 
$\sim$1/4 fm has expanded to the size of $\sim$2 fm at kinetic freezeout. 
With this assumption we find consistency with the theoretical expectations
of the GFTM. We can generate and explain the triggered ridge phenomenon and 
data (see Sec. II), thus implying the ridge is connected to the bubble
substructure. We can predict and obtain very strong evidence for the color 
electric field of the glasma from comparing our multi-particle charged 
particle correlation predictions, and existing experimental correlation 
publications (see Sec. III). We have also predicted correlations which can 
be used to search for evidence for the glasma color magnetic field (Sec. III).

Of course an obvious question that arises is that since a flux tube is an 
isolated system does a PBM bubble, that we assume is the final state of a
flux tube at kinetic freezeout, also act as an isolated system when emitting 
the final state particles? The near side correlations signals since the
original PBM\cite{PBM} have always come virtually entirely from particles
emitted from the same bubble. Thus each bubble has always acted as an 
isolated system similar to the behavior of a flux tube.

\section{The Ridge is formed by the bubbles when a jet trigger is added to the
PBM}

In heavy ion collisions at RHIC there has been observed a phenomenon called the
ridge which has many different 
explanations\cite{Dumitru,Armesto,Romatschke,Shuryak,Nara,Pantuev,Mizukawa,Wong,Hwa}. The ridge is a long range charged particle correlation in $\Delta \eta$ 
(very flat), while the $\Delta \phi$ correlation is approximately jet-like 
(a narrow Gaussian). There also appears with the ridge a jet-like 
charged-particle-pair correlation which is symmetric in 
$\Delta \eta$ and $\Delta \phi$ such that the peak on the
jet-like correlation is at $\Delta \eta$ = 0 and $\Delta \phi$ = 0. The 
$\Delta \phi$ correlation of the jet and the ridge are approximately the same 
and smoothly blend into each other. The ridge correlation is generated when 
one triggers on an intermediate $p_t$ range charged particle and then forms
pairs between that trigger particle and each of all other intermediate charged 
particles with a smaller $p_t$ down to some lower limit. The first case we will
study in this paper is a trigger charged particle between 3.0 to 4.0 GeV/c 
correlated with all other charged particles which have a $p_t$ between 1.1 
GeV/c to 3.0 GeV/c.

In this paper we will investigate whether the PBM can account for the ridge 
once we add a jet trigger to our PBM generator\cite{PBM}. However this trigger 
will also select jets which previously could be neglected because there was
such strong quenching\cite{quench1,quench2,quench3} of jets in central 
collisions. A jet trigger had not been used in the PBM comparison to all
previous data. We use HIJING\cite{hijing} merely to determine the expected 
number of jets to add for our added jet trigger. These jet particles were 
added to our PBM generator. Thus our PBM generator now had HIJING generated 
background particles, bubbles of the PBM and added jet particles from 
HIJING. We have already shown that our final state particles come 
from hadrons at or near the fireball surface. We reduce the 
number of jets by 80\% which corresponds to the estimate that 
only the parton interactions on or near the surface are not 
quenched away, and thus at kinetic freezeout form and emit hadrons which 
enter the detector. This 80\% reduction is consistent with single $\pi^0$ 
suppression observed in Ref.\cite{quench3}. We find for the reduced HIJING 
jets that 4\% of the Au + Au central events (0-5\%) centrality at 
$\sqrt{s_{NN}} =$ 200 have a charged particle with a $p_t$ between 3.0 and 
4.0 GeV/c with at least one other charged particle with its $p_t$ greater than 
1.1 GeV/c coming from the same jet. The addition of the jets to the PBM 
generator provides the appropriate particles which are picked up by the
trigger in order to form a narrow $\Delta \eta$ correlation signal at 0 
which is also a narrow signal in $\Delta \phi$ at 0 (Fig. 10). This narrow jet 
signal is present in the data and is what remains of jets after 80\% are
quenched away. 

We then form two-charged-particle correlations between one-charged-particle 
with a $p_t$ between 3.0 to 4.0 GeV/c and another charged-particle 
whose $p_t$ is greater than 1.1 GeV/c. The results of these 
correlations are shown in Fig. 2. Fig. 2 is the CI correlation for 
the 0-5\% centrality bin with the just above stated $p_t$ 
selections on charge pairs. Since we know in our Monte Carlo which 
particles are emitted from bubbles and thus form the ridge, we can predict the 
shape of the ridge for the above $p_t$ cut by plotting only the correlation 
formed from pairs of particles that are emitted by the same bubble 
(see Fig. 3). The charge pair correlations are virtually all emitted from the 
same bubble. Those charge pair correlations formed from particles originating 
from different bubbles are small contributors which mainly have some effect on 
the away side correlation. Thus Fig. 3 is the ridge signal which is the piece 
of the CI correlation for the 0-5\% centrality of Fig. 2, after removing all 
other pairs except the pairs emitted from the same bubble.

\begin{figure*}[ht] \centerline{\includegraphics[width=0.800\textwidth]
{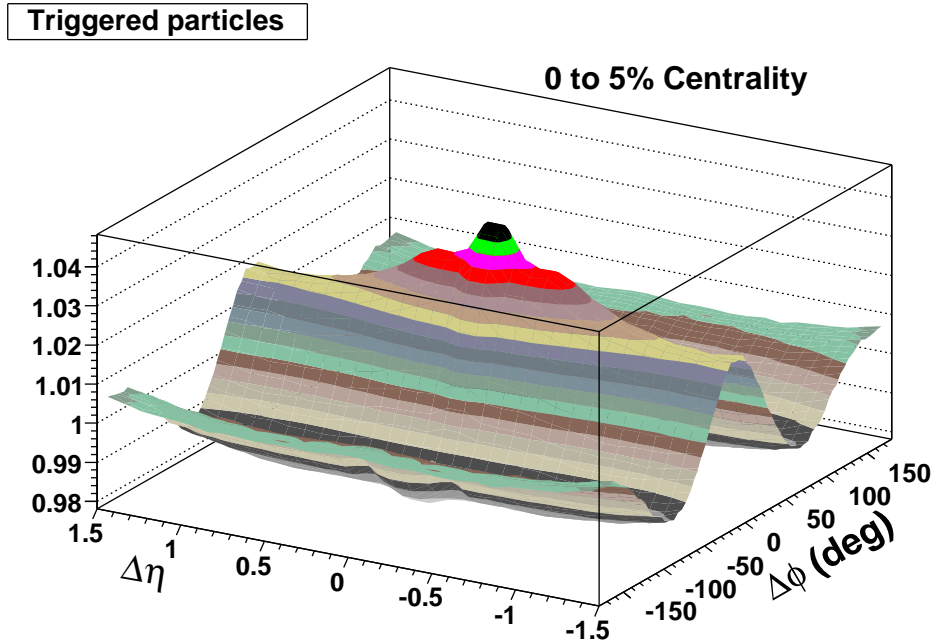}} \caption{The CI correlation for the 0-5\% centrality bin that 
results from requiring one trigger particle $p_t$ above 3 GeV/c and another 
particle $p_t$ above 1.1 GeV/c. It is plotted as a two dimensional 
$\Delta \phi$ vs. $\Delta \eta$ perspective plot.}
\label{fig2}
\end{figure*}
                                                                              
\begin{figure*}[ht] \centerline{\includegraphics[width=0.800\textwidth]
{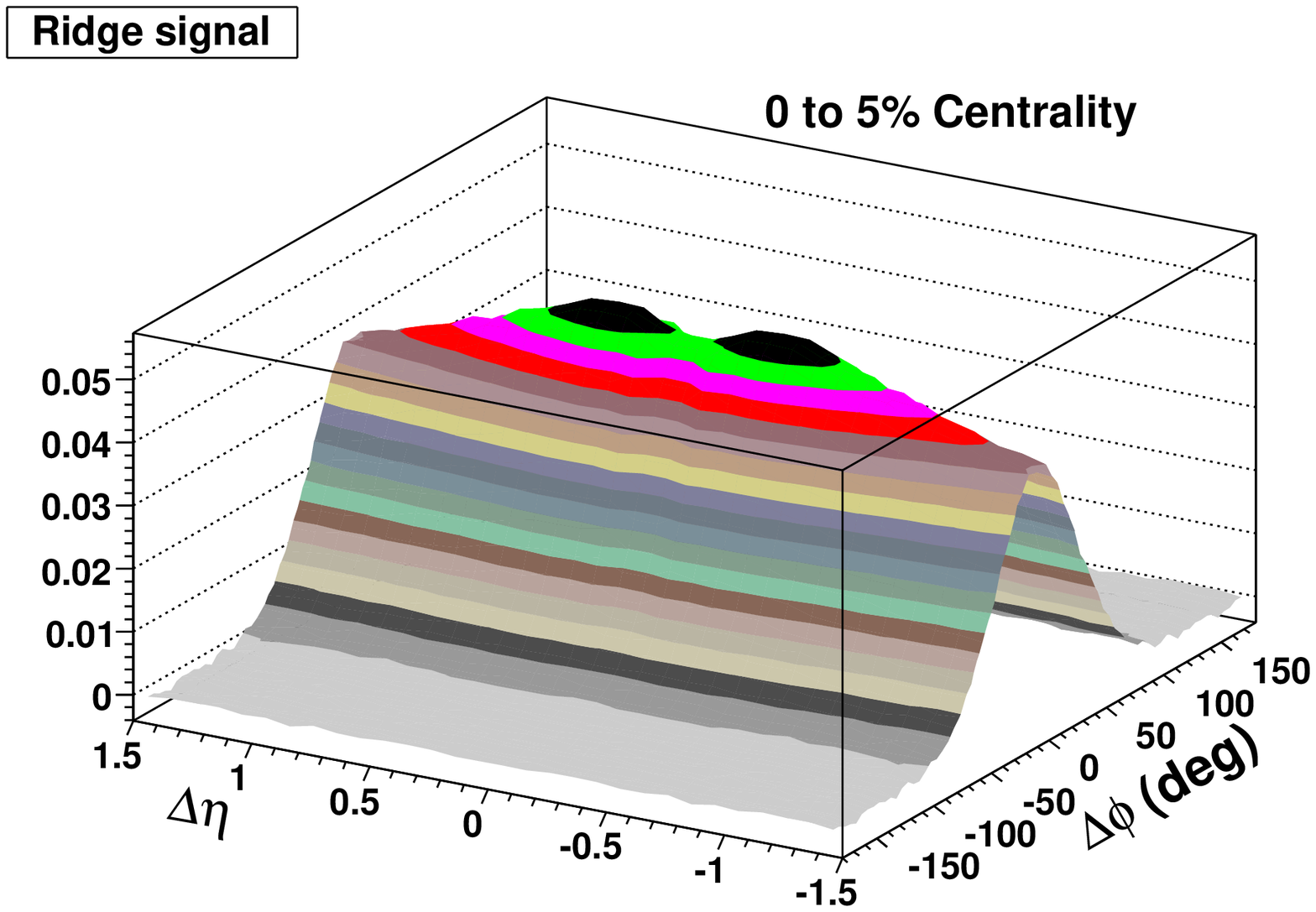}} \caption{The ridge signal is the piece of the CI correlation 
for the 0-5\% centrality of Fig. 2 after removing all other particle pairs 
except the pairs that come from the same bubble. It is plotted as a two 
dimensional $\Delta \phi$ vs. $\Delta \eta$ perspective plot. As explained 
in the text charge pair correlations formed by particles emitted from 
different bubbles are small contributors which mainly have some effect on 
the away side correlation for $\Delta \phi$ greater than about $120^\circ$.}
\label{fig3}
\end{figure*}
                                                                              
\subsection{Parton $p_t$ correlated vs. random}

A very important aspect of the GFTM is the boost that the flux tubes get from 
the radial flow of the blast wave. This boost is the same all along the 
flux tube and only depends on how far away from the center axis of the 
blast wave the flux tube is. In the PBM the partons in the bubble received a
boost in $p_t$ from the radial flow field in the blast wave; that after final
state fragmentation of the bubbles gave the generated particles a 
consistent $p_t$ spectrum with the data. Since the boost from the blast wave 
depended on the position of the bubble in the radial flow field, there should 
be a correlation between partons $p_t$ within a bubble. If one redistributed 
the partons with their $p_t$ boosts uniformly among the bubbles, the over all 
results for the generated correlations without a trigger would be unchanged. 
One should note none of our prior work (i.e.PBM\cite{PBM} or PBME\cite{PBME})
contained a trigger. We have only added a trigger here to treat the
``triggered ridge'' which obviously requires it. In our treatment of the GFTM 
(Sec. III) we have removed the trigger.

Once we require a trigger demanding higher $p_t$ particles, we start 
picking out bubbles which have more radial flow (harder particles). Thus 
correlated particles which pass our $p_t$ selection and trigger come almost 
entirely from the harder bubbles with more radial flow, while the softer 
bubbles which were subject to less radial flow mainly generate low enough $p_t$
particles that become background particles or are lost to the correlation 
analysis due to $p_t$ selection. Figure 2 shows the result of our trigger 
and $p_t$ selection. If one redistributed the partons with their $p_t$ 
radial flow boost among all the bubbles, then all bubbles become equal 
and there is no longer soft and hard bubbles. With this change 
we can generate bubble events. We find that both the non-triggered 
correlation and the $\Delta \phi$ correlation are virtually unchanged 
in the region less than about $120^\circ$ (see Fig. 4). It seems that
we move from the situation where we have a few bubbles with a lot of correlated
particles above our $p_t$ cut to a lot of bubbles that have few correlated 
particles above our cuts. However there is a difference in the triggered 
correlation's away side or $\Delta \phi$ near the $180^\circ$ peak. 

In Ref.\cite{PBM} we discuss this away side effect and attributed this peak 
to momentum conservation. This is however not the whole story. The away 
side peak is generated by the geometry of the bubble ring which requires 
that for every bubble there is another bubble nearly $180^\circ$ away that 
is emitting particles. These symmetry requirements on the geometry 
contribute more to the away side peak especially for the case without a 
trigger or redistributed random partons.

This geometry effect is what causes the away side correlation of elliptic flow 
in a peripheral heavy ion collision. In such a collision there is a higher 
energy density aligned with the reaction plane. On one side of the reaction 
plane there are more particles produced because of increased energy 
density, and because of the geometry of the situation. On the other side of the
reaction plane there are also more particles produced for the same reasons. 
Thus there is a correlation between particles that are near each other on 
one side of the reaction plane and a correlation between particles that are 
on both sides of the reaction plane. Momentum conservation requires 
particle communication by bouncing into each other, while geometry has
no such communication. The right hand does not need to know what the left
hand is doing. However, if physics symmetry makes them do the same thing 
then they can show a correlation.

\subsection{The away side peak}

The away side peak depends on the fragmentation of the away side bubble.
For the triggered case where there is a hard bubble with a strong correlation 
of parton $p_t$ inside the bubble (GFTM like), we will have more correlated 
particles adding to the signal on the near side. On the other hand the 
away side bubble will be on average a softer bubble which will have a lot less
particles passing the $p_t$ cuts and thus have a smaller signal. If we consider
the case of a random parton $p_t$ bubble the particles passing the $p_t$
cuts should be very similar on the near and the away side. In Fig. 4 we
compare the two correlations generated by the correlated and the random 
cases. We compare the $\Delta \phi$ correlation for the two cases (solid is
correlated and open is  random)for each of the five $\Delta \eta$ bins which
cover the entire $\Delta \eta$ range (0.0 to 1.5). The vertical correlation 
scale is not offset and is correct for the largest $\Delta \eta$ bin, 1.2 
$< \Delta \eta <$ 1.5, which is the lowest bin on the figure. As one proceeds 
upwards to the next $\Delta \eta$ bin the correlation is offset by +0.05. 
This is added to the correlation of each subsequent $\Delta \eta$ bin. The 
smallest $\Delta \eta$ bin on top of Fig. 4 has a +0.2 offset. A solid straight
horizontal line shows the offset for each  $\Delta \eta$ bin. Each solid 
straight horizontal line is at 1.0 in correlation strength. We see that
the away side correlation in the $\Delta \phi$ region greater than about
$120^\circ$ is larger for the random case.

\begin{figure*}[ht] \centerline{\includegraphics[width=0.800\textwidth]
{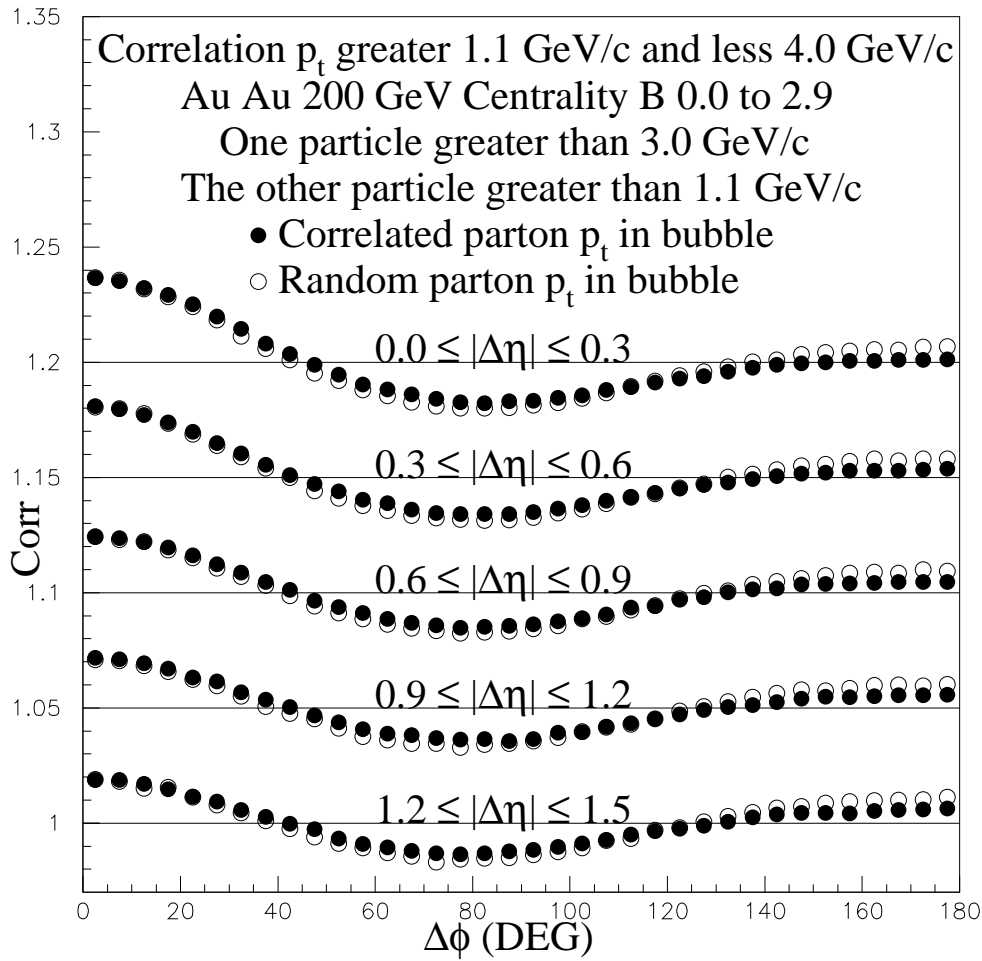}} \caption{The $\Delta \phi$ CI correlation for the 0-5\% 
centrality bin requiring one trigger particle $p_t$ above 3 GeV/c and 
another particle $p_t$ above 1.1 GeV/c. We compare the two correlations 
generated by the correlated and the random cases. We compare the 
$\Delta \phi$ correlation for the two cases (solid is correlated and open 
is random) for each of the five $\Delta \eta$ bins which cover the entire 
$\Delta \eta$ range (0.0 to 1.5). The vertical correlation scale is not 
offset and is correct for the largest $\Delta \eta$ bin, 1.2 $< \Delta \eta <$ 
1.5, which is the lowest bin on the figure. As one proceeds upwards to the 
next $\Delta \eta$ bin the correlation is offset by +0.05. This is added 
to the correlation of each subsequent $\Delta \eta$ bin. The smallest 
$\Delta \eta$ bin on top of Fig. 4 has a +0.2 offset. A solid straight 
horizontal line shows the offset for each  $\Delta \eta$ bin. Each solid 
straight horizontal line is at 1.0 in correlation strength. We see that the 
away side correlation for $\Delta \phi$ greater than about $120^\circ$ is 
larger for the random case.}
\label{fig4}
\end{figure*}
                                                                              
\begin{figure*}[ht] \centerline{\includegraphics[width=0.800\textwidth]
{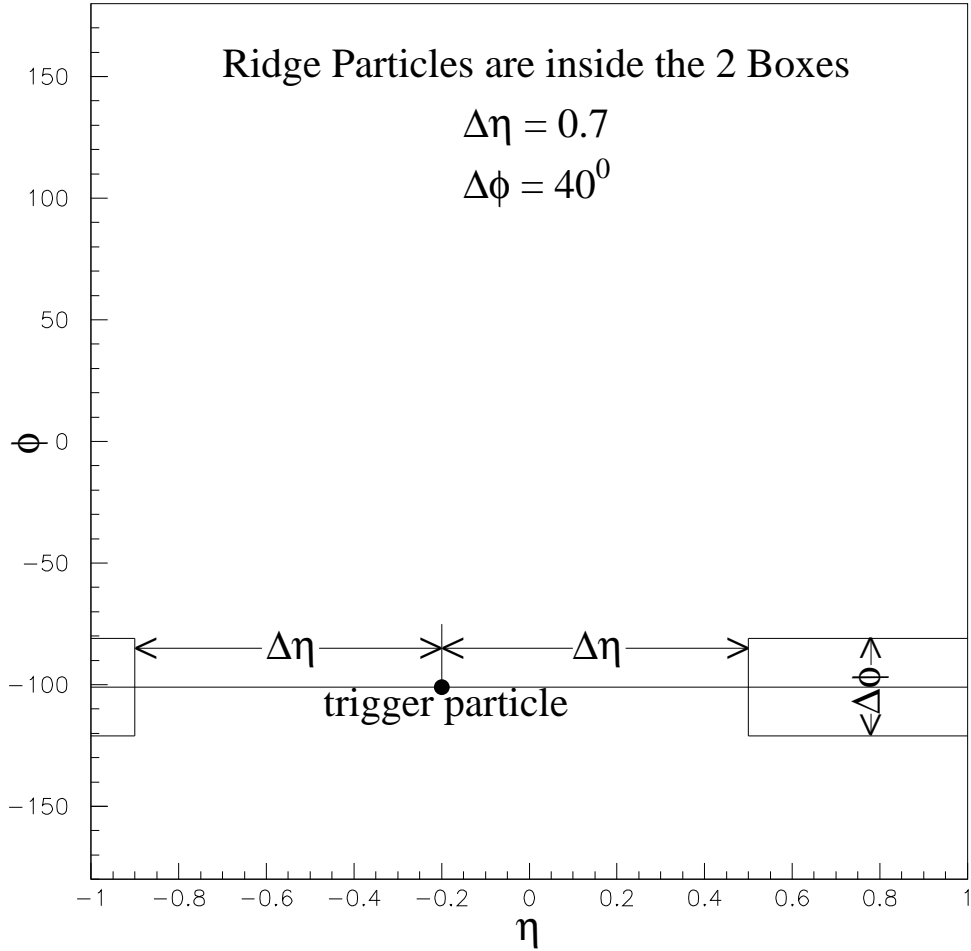}} \caption{In this figure we display the $\phi$ and $\eta$ 
ranges considered for producing ridge particles in the Au + Au central 
collisions. A typical trigger particle ($p_t$ between 3.0 to 4.0 GeV/c) 
implies two associated regions in the considered $\phi$ and $\eta$ ranges 
where the ridge particles lie. The ridge particles we consider are separated 
by 0.7 in $\eta$ ($\Delta \eta$) so that one eliminates particles coming 
from jet production. The bulk of the ridge particles lie within $20^\circ$ of 
the $\phi$ of the trigger particle. The width of the $\phi$ spread is 
$\Delta \phi$ = $40^\circ$.}
\label{fig5}
\end{figure*}
                                                                              
\begin{figure*}[ht] \centerline{\includegraphics[width=0.800\textwidth]
{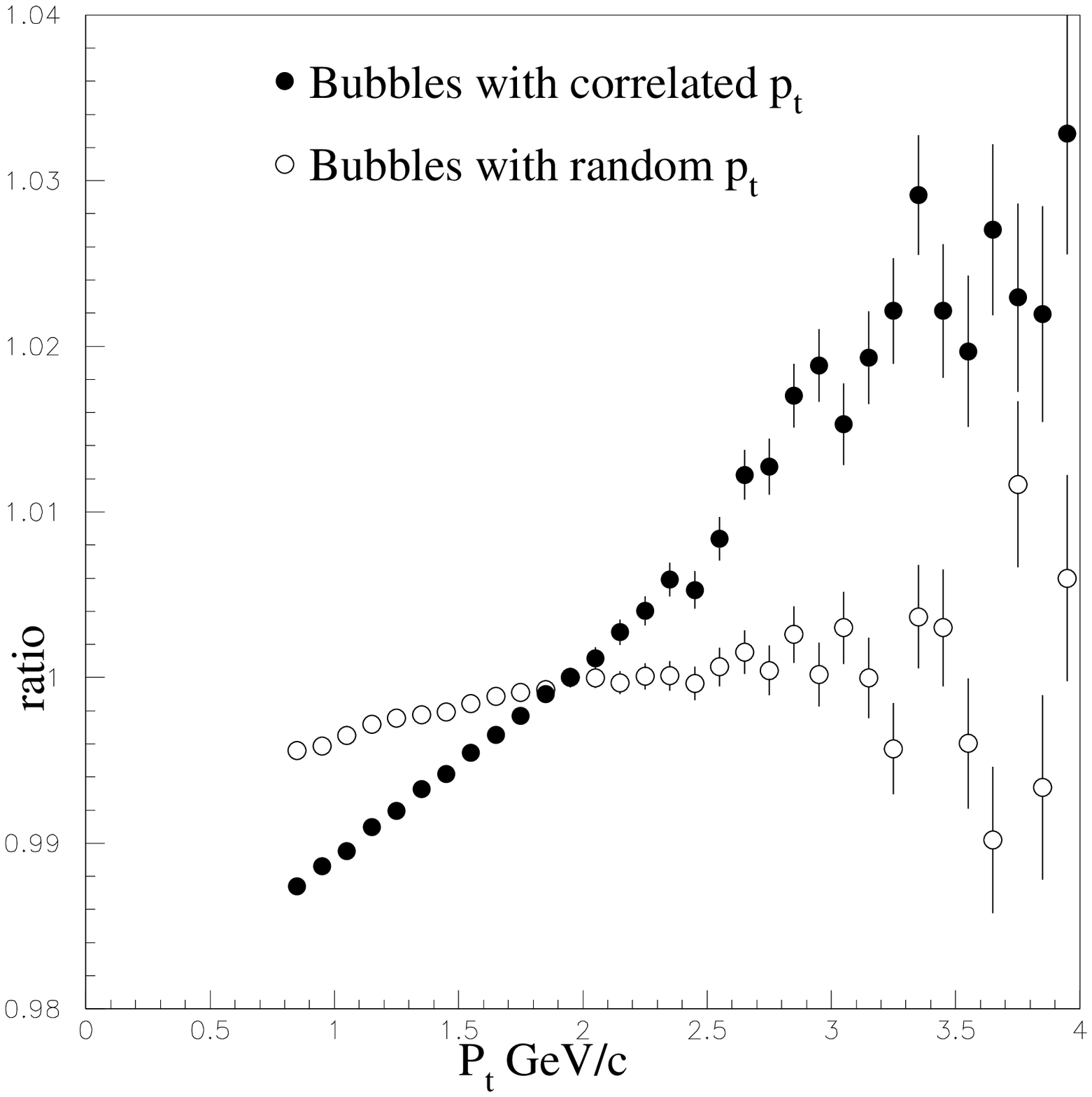}} \caption{On a trigger by trigger basis we accumulate particles 
from the ridge region (see Fig. 5). We then form a ratio the of $p_t$ spectrum 
for the ridge region particles divided by the $p_t$ spectrum for all particles 
in central Au + Au events. We normalize the two $p_t$ spectra to have the same 
counts in the 2.0 GeV/c bin. This ratio or correlation for the bubbles which 
have a correlated $p_t$ among the partons (GFTM like) are the solid points. The
random $p_t$ distribution among partons in the bubble are the open points.}
\label{fig6}
\end{figure*}
                                                                              
\subsection{Predicted transverse momentum dependence of correlation}

Particles which come from bubbles that satisfy the trigger (3.0 $<$ $p_t$ $<$ 
4.0 GeV/c) are harder and have a different $p_t$ distribution compared to 
the case without trigger requirements. In a given triggered event if one could 
select particles that come from the bubble or ridge, one would find that the 
$p_t$ spectrum is harder than the average $p_t$ spectrum. In Fig. 5 we
show how one can by trigger choose particles that are rich with 
ridge particles. We consider a typical trigger particle and the 
associated regions in the available experimental $\phi$ and $\eta$ ranges. 
The ridge particles we consider are separated from the trigger particle
by 0.7 in $\eta$. This is done to eliminate particles coming from jet 
production that could also be associated with the trigger particle. The bulk 
of the ridge particles lie within $20^\circ$ of the $\phi$ of the trigger 
particle. The width of the $\phi$ spread is $\Delta \phi$ = $40^\circ$. 
Therefore on a trigger by trigger basis we accumulate particles from the ridge 
region (see Fig. 5) and form a $p_t$ spectrum of charged-particles-pairs. We 
then form a ratio of the $p_t$ spectrum for the ridge region particles divided 
by the $p_t$ spectrum for all particles in central Au + Au events. This ratio 
is highly dependent on our ridge cut area. We can virtually remove this 
dependency by normalizing the two $p_t$ spectra to have the same counts for 
the 2.0 GeV/c bin. Figure 6 is this ratio which can be considered a correlation
of particles as a function of $p_t$ in the ridge region compared to particles 
from the event in general. In Fig. 6 we form this ratio or correlation for the 
bubbles which have a correlated $p_t$ among the partons (GFTM like). We also 
show the correlation for bubbles which have a random $p_t$ distribution among 
partons in the bubble. For the random case the $p_t$ spectrum does not differ 
much from the average $p_t$ spectrum while there is a big difference for the 
flux tube like case. This correlation could be easily measured in the 
RHIC data.

\subsection{Comparison to data}

Triggered angular correlation data showing the ridge was presented at Quark
Matter 2006\cite{Putschke}. Figure 7 shows the experimental $\Delta \phi$ vs. 
$\Delta \eta$ CI correlation for the 0-10\% central Au + Au collisions at 
$\sqrt{s_{NN}} =$ 200 GeV; requiring one trigger particle $p_t$ between 3 and 
4 GeV/c and an associated particle $p_t$ above 2.0 GeV/c. The yield is
corrected for the finite $\Delta \eta$ pair acceptance. For the PBM generator, 
we then form a two-charged-particle correlation between one charged particle 
with a $p_t$ between 3.0 to 4.0 GeV/c and another charged particle whose $p_t$ 
is greater than 2.0 GeV/c. The results of this correlation is shown in Fig. 
8. Figure 7 shows the corrected pair yield determined in the central data 
whereas Fig. 8 shows the correlation function generated by the PBM which does 
not depend on the number of events analyzed. We can compare the two figures, 
if we realize that the away side ridge has around 420,000 pairs in Fig. 7 
while in Fig. 8 the away side ridge has a correlation of around 0.995. 
If we multiply the correlation scale of Fig. 8 by 422,111 in order to 
achieve the number of pairs seen in Fig. 7, the away side ridge would be at 
420,000 and the peak would be at 465,000. This would make a good agreement 
between the two figures.  

We know in our Monte Carlo which particles come from bubbles and which 
particles form the ridge. The correlation formed by the ridge particles is 
generated almost entirely by particles emitted by the same bubble. We have
shown in all our publications that the same side correlation signals are 
almost entirely formed by particles coming from the same bubble. Thus we can 
predict the shape and the yield of the ridge for the above $p_t$ trigger
selection and lower cut, by plotting only the correlation coming from pairs of 
particles that are emitted by the same bubble (see Fig. 9).

In Ref.\cite{Putschke} it was assumed that the ridge yield was flat across 
the acceptance while in Fig. 9 we see that this is not the case. 
Therefore our ridge yield is approximately 35\% larger than estimated in 
Ref.\cite{Putschke}. Finally we can plot the jet yield that we had put into 
our Monte Carlo. We used HIJING\cite{hijing} to determine the number of 
expected jets, and then reduced the number of jets by 80\%. This assumes that 
only the parton interactions on or near the surface that form hadrons at 
kinetic freezeout are not quenched away and thus enter the detector. This 80\% 
reduction is consistent with single $\pi^0$ suppression observed 
in Ref.\cite{quench3}. This jet yield is plotted in Fig. 10 where we 
subtracted contributions from the bubbles and the background particles 
from Fig. 7.

\section{Strong CP violations or Chern-Simons topological charge}

\subsection{The Source $F \widetilde F$.}

The strong CP problem remains one of the most outstanding puzzles of the
Standard Model. Even though several possible solutions have been put forward
it is not clear why CP invariance is respected by the strong interaction.
It was shown however through a theorem by Vafa-Witten\cite{Witten1} that the
true ground state of QCD cannot break CP. The part of the QCD Lagrangian 
that breaks CP is related to the gluon-gluon interaction term 
$F \widetilde F$. 

\begin{figure*}[ht] \centerline{\includegraphics[width=0.800\textwidth]
{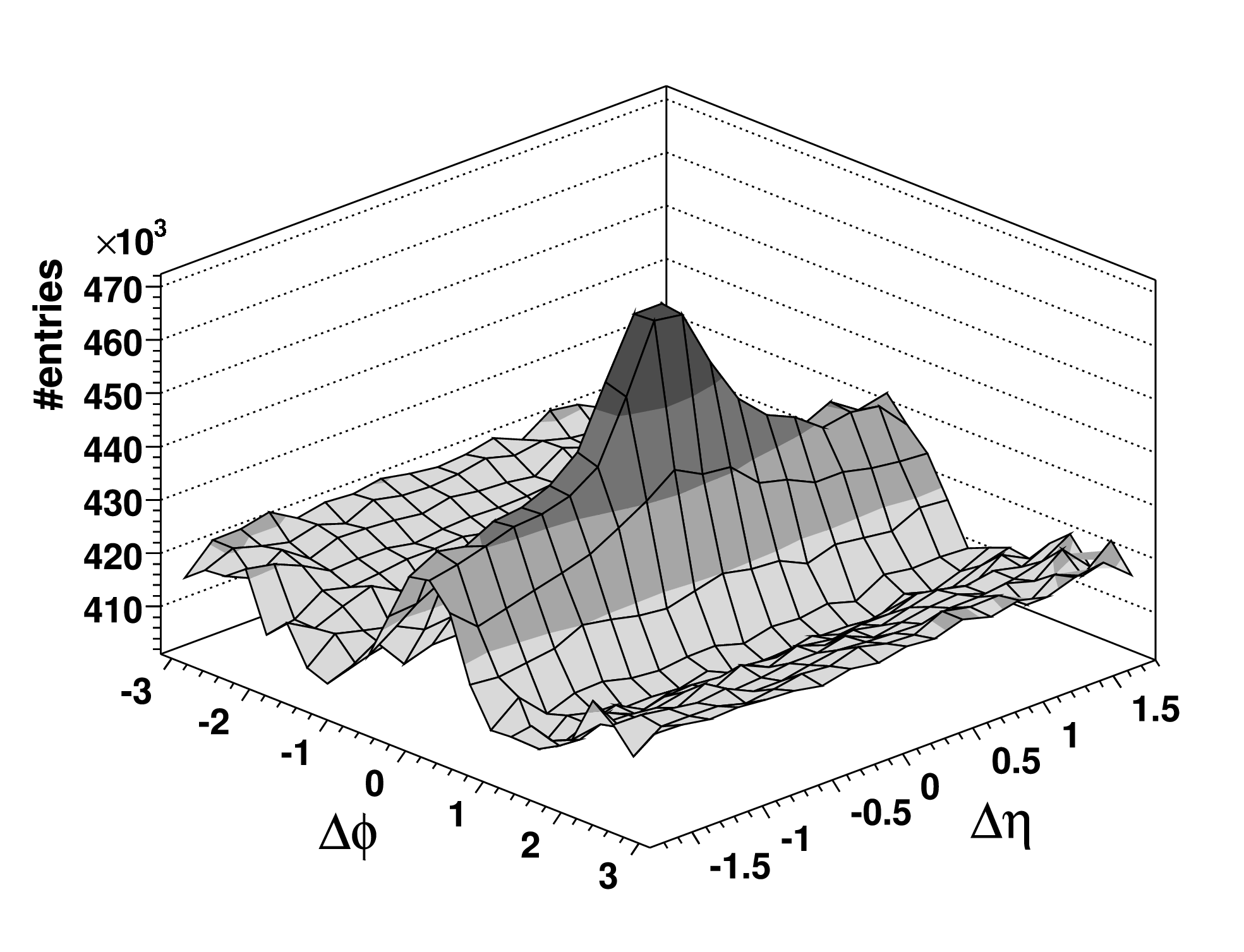}} \caption{Raw $\Delta \phi$ vs. $\Delta \eta$ CI preliminary 
correlation data\cite{Putschke} for the 0-10\% centrality bin for Au + Au 
collisions at $\sqrt{s_{NN}} =$ 200 GeV requiring one trigger particle $p_t$ 
between 3 to 4 GeV/c and an associated particle $p_t$ above 2.0 GeV/c.}
\label{fig7}
\end{figure*}
                                                                              
\begin{figure*}[ht] \centerline{\includegraphics[width=0.800\textwidth]
{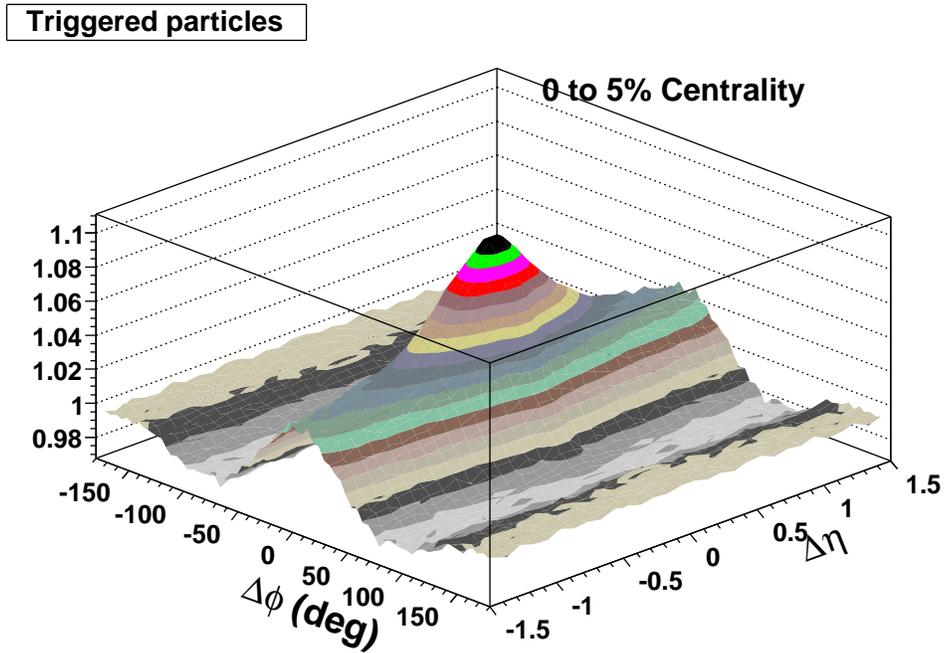}} \caption{The PBM generated CI correlation for the 0-5\% 
centrality bin requiring one trigger particle $p_t$ above 3 GeV/c and another 
particle $p_t$ above 2.0 GeV/c plotted as a two dimensional $\Delta \phi$ vs. 
$\Delta \eta$ perspective plot. The trigger requirements on this figure are 
the same as those on the experimental data in Fig. 7.}
\label{fig8}
\end{figure*}
                                                                              
\begin{figure*}[ht] \centerline{\includegraphics[width=0.800\textwidth]
{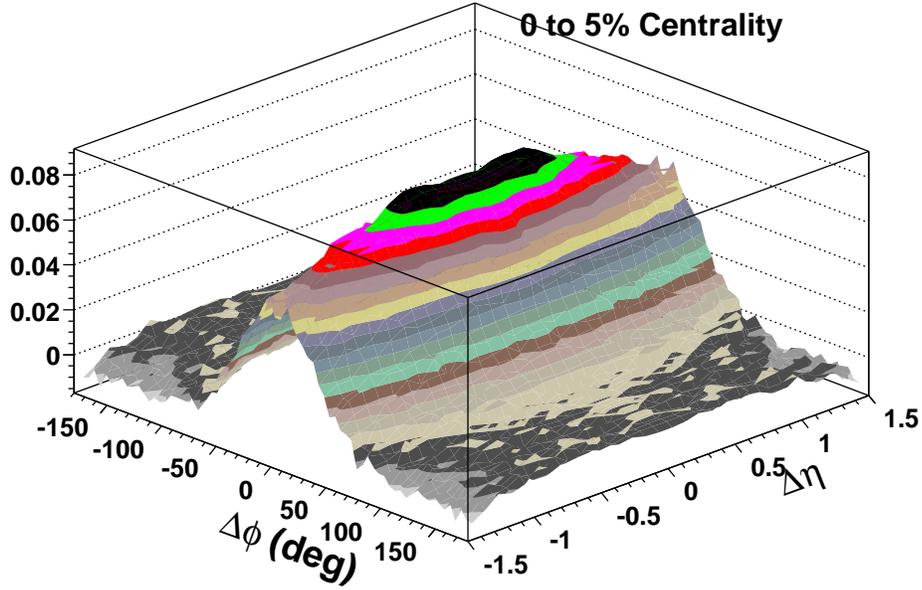}} \caption{The ridge signal is the piece of the CI correlation 
for the 0-5\% centrality of Fig. 8 after removing all other particle pairs 
except the pairs that come from the same bubble. It is plotted as a two 
dimensional $\Delta \phi$ vs. $\Delta \eta$ perspective plot.}
\label{fig9}
\end{figure*}
                                                                              
\begin{figure*}[ht] \centerline{\includegraphics[width=0.800\textwidth]
{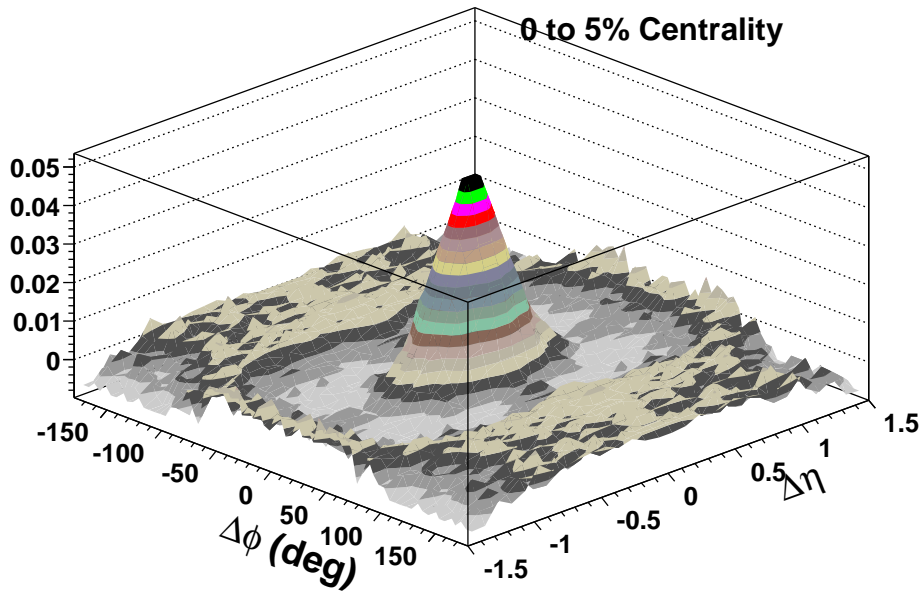}} \caption{The jet signal is left in the CI correlation after 
the contributions from the background and all the bubble particles are removed 
from the 0-5\% centrality (with trigger requirements) of Fig. 8. It is plotted 
as a two dimensional $\Delta \phi$ vs. $\Delta \eta$ perspective plot.}
\label{fig10}
\end{figure*}
                                                                              
\clearpage

The part of $F \widetilde F$ related to CP violations 
can be separated into a separate term which then can be varied by 
multiplying this term by a parameter called $\theta$. For the true 
QCD ground state $\theta$ = 0 (Vafa-Witten theorem). In the vicinity 
of the deconfined QCD vacuum metastable domains\cite{Tytgat} 
$\theta$ non-zero could exist and not contradict the Vafa-Witten theorem 
since these are not the QCD ground state. The metastable 
domains CP phenomenon would manifest itself in specific correlations of 
pion momenta\cite{Pisarski,Tytgat}.

\subsection{Pionic measures of CP violation.}

The glasma flux tube model (GFTM)\cite{Dumitru} considers the 
wavefunctions of the incoming projectiles, form sheets of CGC\cite{CGC}
at high energies that collide, interact, and evolve into high intensity color 
electric and magnetic fields. This collection of primordial fields is 
the Glasma\cite{Lappi,Gelis}. Initially the Glasma is composed of 
only rapidity independent longitudinal (along the beam axis) 
color electric and magnetic fields. These longitudinal 
color electric and magnetic fields generate topological 
Chern-Simons charge\cite{Simons} through the $F \widetilde F$ 
term and becomes a source of CP violation. How much of these 
longitudinal color electric and magnetic fields are still present in the 
surface flux tubes when they have been pushed by the blast wave is a 
speculation of this paper for measuring strong CP violation? The color 
electric field which points along the flux tube axis causes an up quark to be
accelerated in one direction along the beam axis, while the anti-up quark 
is accelerated in the other direction. So when a pair of quarks and 
anti-quarks are formed they separate along the beam axis leading to 
a separated $\pi^+$ $\pi^-$ pair along this axis. The color magnetic field 
which also points along the flux tube axis (which is parallel 
to the beam axis) causes an up quark to rotate around the flux tube axis 
in one direction, while the anti-up quark is rotated in the other direction. 
So when a pair of quarks and anti-quarks are formed they will pickup or lose 
transverse momentum. These changes in $p_t$ will be transmitted to the 
$\pi^+$ $\pi^-$ pairs.

\it It is important to note that the CP violating asymmetries in 
$\pi^+$ and $\pi^-$ momenta arise through the 
Witten-Wess-Zumino\rm\cite{Witten2,Wess} \it term. The quarks and anti quarks 
which, later form the $\pi^+$ and $\pi^-$, directly respond to the color 
electric and color magnetic fields and receive their boosts at the quark and
anti quark level before they are pions. These boosts are transmitted to the
pions at hadronization. Thus no external magnetic field is required in the
methodology followed in this paper. The following references demonstrate 
this:\rm\cite{2000,Pisarski,Tytgat,Finch}.

To represent the color electric field effect we assume as the first step 
for the results being shown in this paper; that we generate bubbles which have 
an added boost of 100 MeV/c to the quarks in the longitudinal momentum which 
represents the color electric effect. The $\pi^+$ and the $\pi^-$ which form a 
pair are boosted in opposite directions along the beam axis. In a given bubble 
all the boosts are the same but vary in direction from bubble to bubble. For
the color magnetic field effect, we give 100 MeV/c boosts to the transverse 
momentum in an opposite way to the $\pi^+$ and the $\pi^-$ which form a pair. 
For each pair on one side of a bubble one of the charged pions boost is 
increased and the other is decreased. While on the other side of the bubble 
the pion for which the boost was increased is now decreased and the pion 
which was decreased is now increased by the 100 MeV/c. All pairs for a 
given bubble are treated in the same way however each bubble is random 
on the sign of the pion which is chosen to be boosted on a given side. 
This addition to our model is used in the simulations of the following 
subsections.

\subsection{Color electric field pionic measure.}

Above we saw that pairs of positive and negative pions should show a charge 
separation along the beam axis due to a boost in longitudinal momentum
caused by the color electric field. A measure of this separation should be a 
difference in the pseudorapidity ($\Delta \eta$) of the opposite sign pairs. 
This $\Delta \eta$ measure has a well defined sign since we defined this 
difference measuring from the $\pi^-$ to the $\pi^+$. In order to form a 
correlation we must pick two pairs for comparison. The pairs have to come from 
the same bubble (a final state of an expanded flux tube) since we have
shown by investigating the events generated by the model that pairs 
originating from different bubbles will not show this correlation. 
Therefore we require the $\pi^+$ and the $\pi^-$ differ by $20^\circ$ 
or less in $\phi$. Let us call the first pair $\Delta \eta_1$. The 
next pair ($\Delta \eta_2$ having the same $\phi$ requirement) has to also lie 
inside the same bubble to show this correlation. Thus we require that there is 
only $10^\circ$ between the average $\phi$ of each bubble. This implies that 
at most in $\phi$ no two pions can differ by more than $30^\circ$. In 
Fig. 11 we show two pairs which would fall into the above cuts. 
$\Delta \eta_1$ and $\Delta \eta_2$ are positive in Fig. 11. However if we 
would interchange the $\pi^+$ and $\pi^-$ on either pair the value of its 
$\Delta \eta$ would change sign. Finally the mean value shown on Fig. 11 
is the mid-point between the $\pi^+$ and the $\pi^-$ where one really uses 
the vector sum of the $\pi^+$ and the $\pi^-$ which moves this point toward 
the harder pion.

Considering the above cuts we defined a correlation function where we combine 
pairs each having a $\Delta \eta$. Our variable is related to the sum of the 
absolute values of the individual $\Delta \eta$'s 
($\vert \Delta \eta_1 \vert +\vert \Delta \eta_2 \vert$). We assign a sign
to this sum such that if the sign of the individual $\Delta \eta$'s are
the same it is a plus sign, while if they are different it is a minus sign.
For the flux tube the color electric field extends over a large pseudorapidity
range therefore let us consider the separation of pairs 
$\vert \Delta \eta \vert$ greater than 0.9. For the numerator of the
correlation function we consider all combinations of unique pairs
(sign ($\vert \Delta \eta_1 \vert +\vert \Delta \eta_2 \vert$)) from a given 
central Au + Au event divided by a mixed event denominator created from pairs
in different events. We determine the rescale of the mixed event denominator by
considering the number of pairs of pairs for the case $\vert \Delta \phi \vert$
lying between $50^\circ$ and $60^\circ$ for events and mixed events so that the
overall ratio of this sample numerator to denominator is 1. By picking 
$50^\circ$ $<$ $\vert \Delta \phi \vert$ $<$ $60^\circ$ we make sure we are
not choosing pairs from the same bubble. For a simpler notation let 
(sign ($\vert \Delta \eta_1 \vert +\vert \Delta \eta_2 \vert$)) =
$\Delta \eta_1 + \Delta \eta_2$ which varies from -4 to +4 since we have an
over all $\eta$ acceptance -1 to +1 (for the STAR TPC detector for which we 
calculated). The value being near $\pm$ 4 can happen when one has a hard 
pion with $p_t$ of 4 GeV/c (upper cut) at $\eta$ = 1 with a soft pion $p_t$ 
of 0.8 GeV/c (lower cut) at $\eta$ = -1 combined with another pair; 
a hard pion with $p_t$ of 4 GeV/c at $\eta$ = -1 with a soft pion with
$p_t$ of 0.8 GeV/c at $\eta$ = 1.

\begin{figure*}[ht] \centerline{\includegraphics[width=0.800\textwidth]
{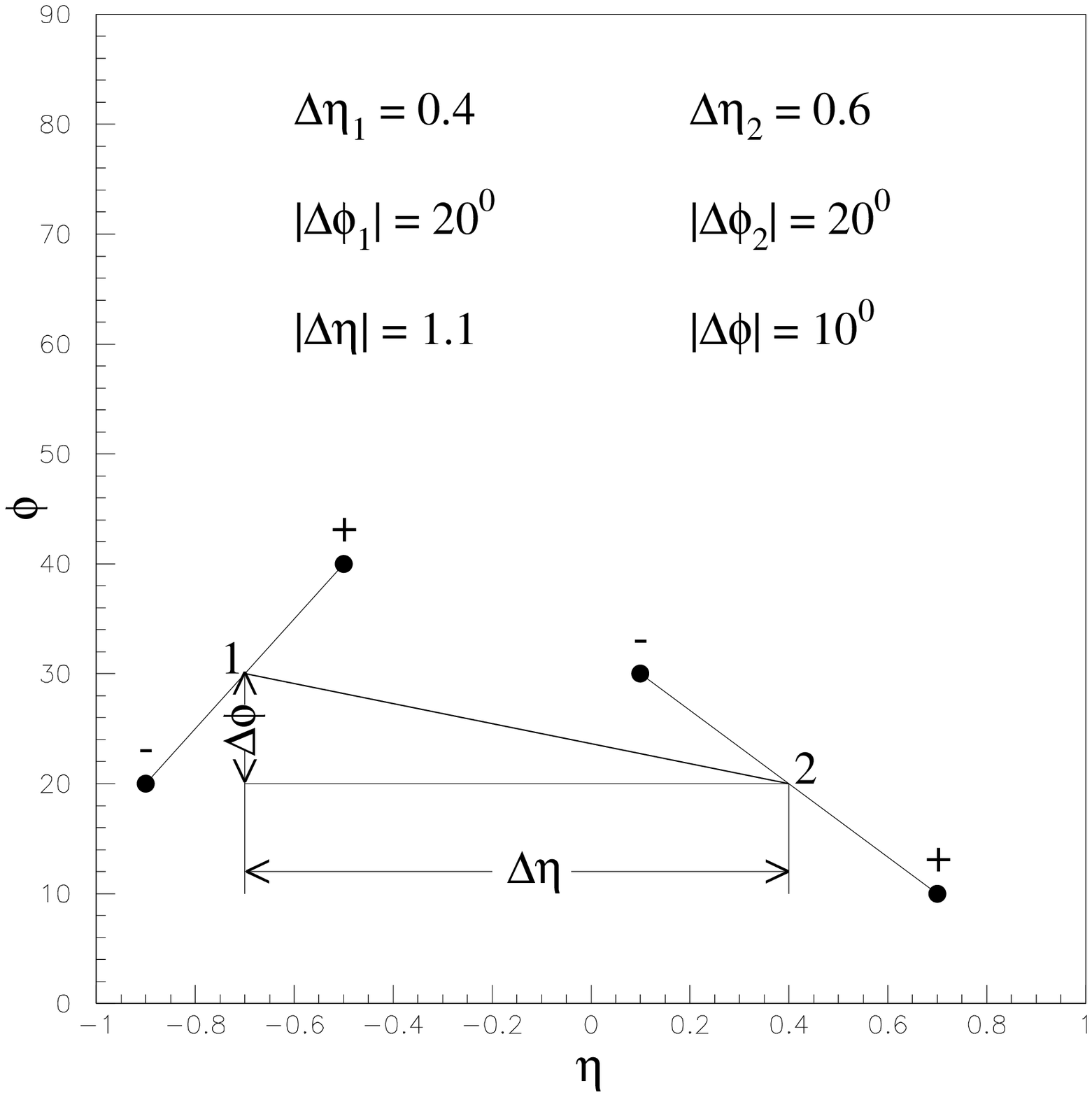}} \caption{Pairs of pairs selected for forming a correlation due 
to the boost in longitudinal momentum caused by the color electric field. The 
largest separation we allow in $\phi$ for plus minus pair 1 is $\Delta \phi_1$ 
= $20^0$. This assures the pair is in the same bubble. The same is true for 
pair 2 so that it will also be contained in this same bubble. The mid-point for
pair 1 and 2 represents the vector sum of pair 1 and 2
which moves toward the harder particle when the momenta differ. These
mid-points can not be separated by more than $10^0$ in $\Delta \phi$ in 
order to keep all four particles inside the same bubble since the correlation
function is almost entirely generated within the same bubble. The 
$\Delta \eta$ measure is the angle between the vector sum 1 compared to the 
vector sum 2 along the beam axis (for this case $\vert \Delta \eta \vert$ = 
1.1). The positive sign for $\Delta \eta_1$ comes from the fact that one moves 
in a positive $\eta$ direction from negative to positive. The same is true for 
$\Delta \eta_2$. If we would interchange the charge of the particles of the 
pairs the sign would change.}
\label{fig11}
\end{figure*}
                                                                              
\begin{figure*}[ht] \centerline{\includegraphics[width=0.800\textwidth]
{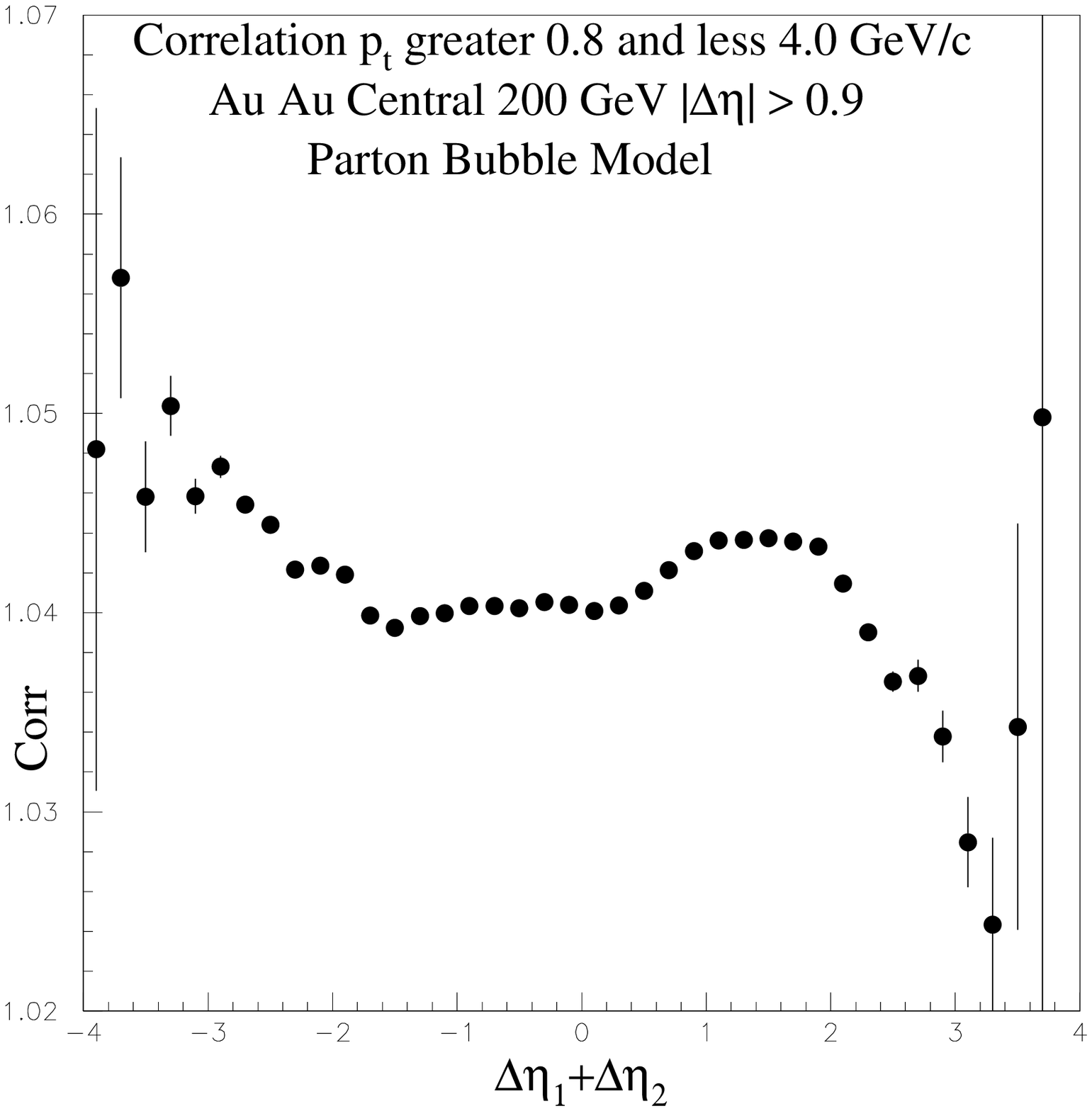}} \caption{ Correlation function of pairs formed to exhibit the 
effects of longitudinal momentum boosts by the color electric field as defined 
in the text. This correlation shows that there are more aligned pairs 
(correlation is larger by $\sim$0.5\% between 1.0 $<$ $\Delta \eta_1 + 
\Delta \eta_2$ $<$ 2.0 compared to between -2.0 $<$ 
$\Delta \eta_1 + \Delta \eta_2$ $<$ -1.0). $\Delta \eta_1 + \Delta \eta_2$ 
is equal to $\vert \Delta \eta_1 \vert +\vert \Delta \eta_2 \vert$.
As explained in the text this means there are more pairs of pairs aligned 
in the same direction compared to pairs not aligned as predicted by the 
color electric field. The signs and more detail are also explained in the 
text.}
\label{fig12}
\end{figure*}
                                                                              
In Fig. 12 we show the correlation function of opposite sign 
charge-particle-pairs paired and binned by the variable 
$\Delta \eta_1 + \Delta \eta_2$ with a cut $\vert \Delta \eta \vert$ 
greater than 0.9 between the vector sums
of the two pairs.  The events are generated by the PBM\cite{PBM} and are 
charged particles of 0.8 $<$ $p_t$ $<$ 4.0 GeV/c, and $\vert \eta \vert$ $<$ 
1, from Au + Au collisions at $\sqrt{s_{NN}} =$ 200 GeV. Since we select pairs 
of pairs which are near each other in $\phi$ they all together pick up the
bubble correlation and thus these pairs of pairs over all show about a
4\% correlation. In the 1.0 $<$ $\Delta \eta_1 + \Delta \eta_2$ $<$ 2.0
region the correlation is 0.5\% larger than the
-2.0 $<$ $\Delta \eta_1 + \Delta \eta_2$ $<$ -1.0 region. This means there
are more pairs of pairs aligned in the same direction compared to pairs of
pairs not aligned. This alignment is what is predicted by the color electric
field effect presented above. In fact if one has plus minus pairs all
aligned in the same direction and spread across a pseudorapidity range,
locally at any place in the pseudorapidity range one would observe
an increase of unlike sign charge pairs compared to like sign charge pairs.
At small $\Delta \eta$ unlike sign charge pairs are much larger than like
sign charge pairs in both the PBM and the data which agree. See Figs. 10, 11,
and 14 of Ref.\cite{PBM}. Figure 11 of Ref.\cite{PBM} compares the total 
correlation for unlike-sign charge pairs and like-sign charge pairs in the 
precision STAR central production experiment for Au + Au central collisions 
(0-10\% centrality) at $\sqrt{s_{NN}}$=200 GeV, in the transverse momentum 
range 0.8 $<$ $p_t$ $<$ 2.0 GeV/c\cite{centralproduction}. The unlike-sign 
charge pairs are clearly larger in the region near $\Delta \phi$ = 
$\Delta \eta$ = 0.0. The increased correlation of the unlike-sign pairs is 
0.8\% $\pm$ 0.002\%. Figure 10 of Ref.\cite{PBM} shows that the PBM fit to 
these data gives the same results. Figure 14 of Ref.\cite{PBM} shows
that the CD = unlike-sign charge pairs minus like-sign charge pairs is positive
for the experimental analysis. Therefore the unlike-sign charge pairs are
considerably larger than the like-sign charge pairs. In fact
this effect is so large and the alignment is so great that when one adds
the unlike and like sign charge pairs correlations together there is still
a dip at small $\Delta \eta$ and $\Delta \phi$ see Fig. 1 and Fig. 3 of the
present paper. The statistical significance of this dip in the two high 
precision experiments done independently from different data sets gathered 2 
years apart\cite{centralproduction,centralitydependence} is huge. It would 
require a fluctuation of $14\sigma$ to remove the dip. This dip is also not 
due to any systematic error since both of the just cited precision experiments 
carefully investigated that possibility; and found no evidence to challenge 
the reality of this dip. This highly significant dip ($\sim$$14\sigma$) means 
that like-sign pairs are removed as one approaches the region $\Delta \eta$ = 
$\Delta \phi$ = 0.0. Thus this is very strong evidence for the predicted 
effect of the color electric field.

\subsection{Color magnetic pionic measure.}
  
After considering the color electric effect we turn to the color magnetic
effect which causes up quarks to rotate around the flux tube axis in one 
direction, while the anti-up quarks rotate in the other direction. So when 
a pair of quarks and anti-quarks are formed they will pickup or lose transverse
momentum. These changes in $p_t$ will be transmitted to the $\pi^+$ and 
$\pi^-$ pairs. \it It was previously shown in Sec. III A that the CP violating
asymmetries in the $\pi^+$ and the $\pi^-$ momenta arise through the 
Witten-Wess-Zumino term. The quarks and anti quarks which, later form the
$\pi^+$ and the $\pi^-$, directly respond to the color electric and color 
magnetic fields and receive their boosts at the quark and anti quark level
before they hadronize into pions. These boosts are transmitted to the $\pi^+$ 
and the $\pi^-$.

\rm In order to observe these differential $p_t$ changes one must
select pairs on one side of the bubble in $\phi$ and compare to other pairs
on the other side of the same bubble which would lie around $40^\circ$ to 
$48^\circ$ away in $\phi$. We defined a pair as a plus particle and minus 
particle with an opening angle ($\theta$) of $16^\circ$ or less. We are 
also interested in pairs that are directly on the other side of the 
bubble. We require they are near in pseudorapidity ($\Delta \eta$ $<$ 0.2). 
The above requirements constrain the four charged particles comprising both
pairs to be contained in the same bubble and be close to being directly
on opposite sides of the bubble (a final state expanded flux tube).
The difference in $p_t$ changes due to and predicted by the color magnetic 
effect should give the $\pi^+$ on one side of the bubble an increased 
$p_t$ and a decreased $p_t$ on the other side, while for the 
$\pi^-$ it will be the other way around. This will lead to an 
anti-alignment between pairs. In Fig. 13 we show two pairs which 
would fall into the above cuts. Both pairs are at the limit 
of the opening angle cut $\theta_1$ and $\theta_2$ equal $16^\circ$. 
The $p_t$ of the plus particle for pair number 1 is 1.14 GeV/c, 
while the minus particle is 1.39 GeV/c. Thus $\Delta P_{t1}$ 
is equal to -0.25 GeV/c. The $p_t$ of the plus particle for 
pair number 2 is 1.31 GeV/c, the minus particle is 0.91 GeV/c 
and $\Delta P_{t2}$ is equal to 0.40 GeV/c. Finally the 
mean value shown on Fig. 13 is the mid-point between the $\pi^+$
and the $\pi^-$ where one really uses the vector sum of the $\pi^+$
and the $\pi^-$ which moves this point toward the harder pion.

\begin{figure*}[ht] \centerline{\includegraphics[width=0.800\textwidth]
{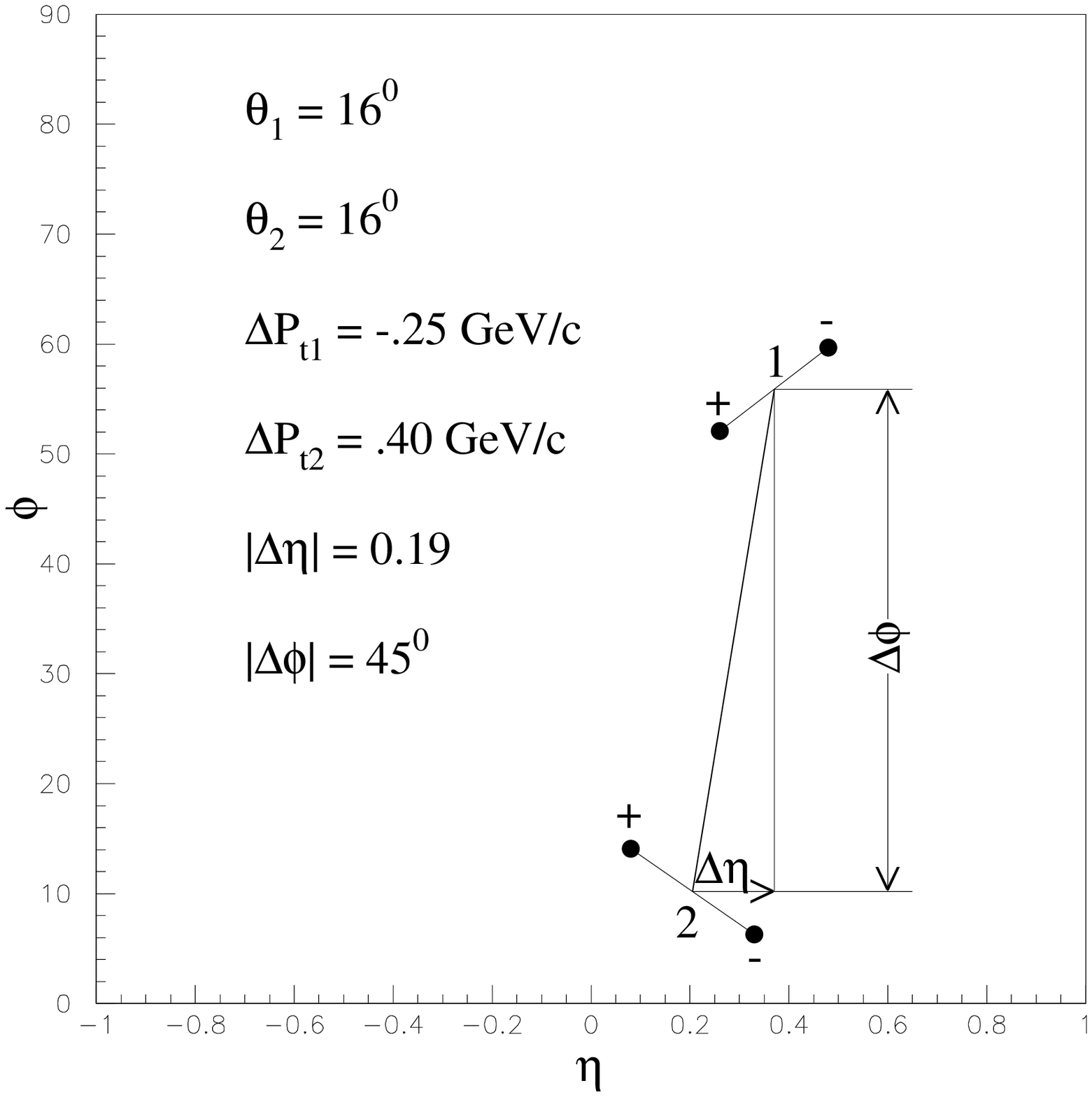}} \caption{Pairs of pairs selected for forming a correlation 
exhibiting the effect of changes in $p_t$ due to the color magnetic field. The 
largest opening angle $\theta$ for plus minus pairs is $16^\circ$ or less. This
opening angle assures that each pair has a high probability that it arises
from a  quark anti-quark pair. In this figure we have picked two pairs
at this limit ($\theta_1$ = $16^\circ$ and $\theta_2$ = $16^\circ$).
The mid-point for pair 1 and 2 represents the vector sum of pair 1 and 2
which moves toward the harder particle when the momenta differ. These
mid-points are chosen to have $40^\circ$ $<$ $\vert \Delta \phi \vert$ $<$
$48^\circ$ in order for the pairs to be on opposite sides of the bubble. 
Since we are interested in pairs directly across the bubble we make the 
$\Delta \eta$ separation be no more than 0.2. The difference in $p_t$
for pair 1 is $\Delta P_{t1}$ = -0.25 GeV/c, while the difference in $p_t$
for pair 2 is $\Delta P_{t2}$ = 0.40 GeV/c. The minus sign for 1 follows
from the fact that the plus particle has 1.14 GeV/c and the minus particle
has 1.39 GeV/c. The plus sign for 2 follows from the fact that the plus 
particle has 1.31 GeV/c and the minus particle has 0.91 GeV/c.}
\label{fig13}
\end{figure*}
                                                                              
Considering the above cuts we defined a correlation function where we combine 
pairs each having a $\Delta P_t$. Our variable is related to the sum of the 
absolute values of the individual $\Delta P_t$'s 
($\vert \Delta P_{t1} \vert +\vert \Delta P_{t2} \vert$). We assign a sign
to this sum such that if the sign of the individual $\Delta P_t$'s are
the same it is a plus sign, while if they are different it is a minus sign.
For the flux tube the color magnetic field extends over a large pseudorapidity
range where quarks and anti-quarks rotate around the flux tube axis, therefore 
we want to sample pairs at the different sides of the tube making a separation
in $\phi$ ($\Delta \phi$) between $40^\circ$ to $48^\circ$. We are interested
in sampling the pairs on the other side so we require the separation in $\eta$
($\Delta \eta$) be 0.2 or less. For the numerator of the
correlation function we consider all combinations of unique pairs
(sign ($\vert \Delta P_{t1} \vert +\vert \Delta P_{t2} \vert$)) from a given 
central Au + Au event divided by a mixed event denominator created from pairs
in different events. We determine the rescale of the mixed event denominator by
considering the number of pairs of pairs for the case $\vert \Delta \eta \vert$
lying between 1.2 and 1.5 plus any value of $\vert \Delta \phi \vert$ for 
events and mixed events so that the overall ratio of this sample numerator to 
denominator is 1. By picking this $\Delta \eta$ bin for all
$\vert \Delta \phi \vert$ we have around the same pair count as the signal
cut with the $\Delta \phi$ correlation of the bubbles being washed out.
For a simpler notation let 
(sign ($\vert \Delta P_{t1} \vert +\vert \Delta P_{t2} \vert$)) =
$\Delta P_{t1} + \Delta P_{t2}$ which we plot in the range from -4 to +4 since 
we have an over all $p_t$ range 0.8 to 4.0 GeV/c. Thus the maximum magnitude 
of $\Delta P_t$'s is 3.2 GeV/c which makes  $\Delta P_{t1} + \Delta P_{t2}$ 
have a range of $\pm$6.4. However the larger values near these range limits
occur very rarely.
 
\begin{figure*}[ht] \centerline{\includegraphics[width=0.800\textwidth]
{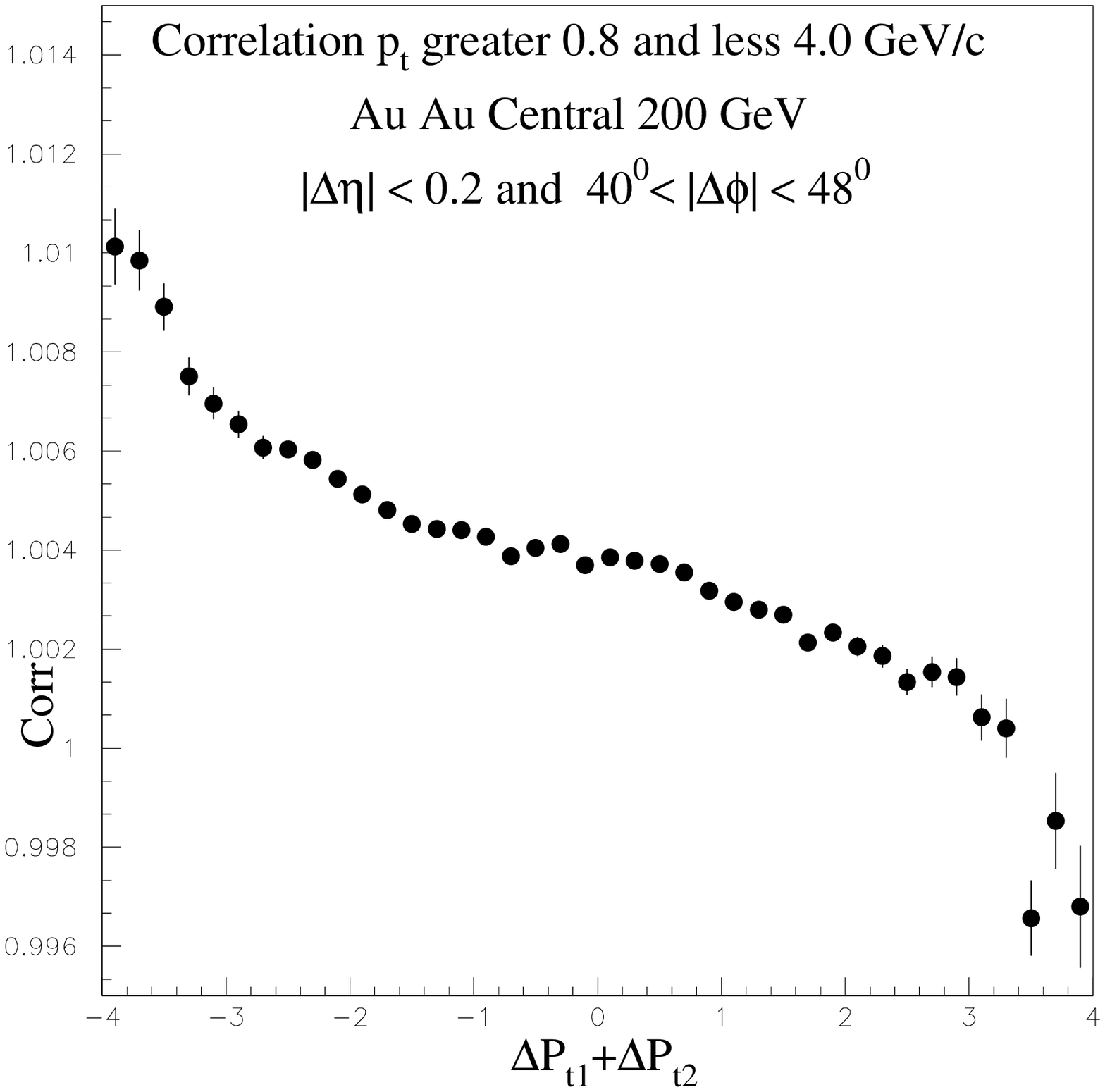}} \caption{ Correlation function of pairs of pairs formed for 
exhibiting the effect of changes in $p_t$ due to the color magnetic field as 
defined in the text. This correlation shows that there are more anti-aligned 
pairs. The correlation is larger by $\sim$1.2\%  for 
$\Delta P_{t1} + \Delta P_{t2}$ = -4.0 compared to 
$\Delta P_{t1} + \Delta P_{t2}$ = 4.0. 
$\Delta P_{t1} + \Delta P_{t2}$ is equal to 
$\vert \Delta P_{t1} \vert +\vert \Delta P_{t2} \vert$ 
where the sign is also explained in the text.}
\label{fig14}
\end{figure*}
                                                                              
In Fig. 14 we show the correlation function of opposite sign
charged-particle-pairs paired and binned by the variable 
$\Delta P_{t1} + \Delta P_{t2}$ with a cut $\vert \Delta \eta \vert$ 
less than 0.2 between the vector sums
of the two pairs, and with $40^\circ$ $<$ $\vert \Delta \phi \vert$ $<$ 
$48^\circ$. The events are generated by the PBM\cite{PBM} and are charged 
particles of 0.8 $<$ $p_t$ $<$ 4.0 GeV/c, with $\vert \eta \vert$ $ <$ 1, from 
Au + Au collisions at $\sqrt{s_{NN}} =$ 200 GeV. Since we select pairs of pairs
which are near each other in $\phi$ ($40^\circ$ to $48^\circ$) they all 
together pick up the bubble correlation and thus these pairs of pairs over all 
show about a 0.4\% correlation. In the -4.0 $<$ $\Delta P_{t1} + \Delta P_{t2}$
$<$ -1.0 region the correlation increases from 0.4\% to 1\%, while in the
1.0 $<$ $\Delta P_{t1} + \Delta P_{t2}$ $<$ 4.0 region the correlation
decreases from 0.4\% to -0.2\%. This means the pairs of pairs are anti-aligned 
at a higher rate than aligned. This anti-alignment is what is predicted by 
the color magnetic field effect presented above. In fact the anti-alignment 
increases with $p_t$ as the ratio bubble particle to background increases.
However these predicted color magnetic effects have not been searched for yet
and therefore there is no experimental evidence for them. Experimental 
investigations of our presented four charged particle correlations provide a
method of obtaining experimental confirmation of the anti-alignment caused 
by the color magnetic field.

\section{Summary and Discussion}

In this article we have made a direct connection between our Parton Bubble
Model (PBM\cite{PBM}) and the Glasma Flux Tube Model (GFTM)\cite{Dumitru}). 
In the GFTM a flux tube is formed right after the initial collision 
of the Au + Au system. This flux tube extends over many units of 
pseudorapidity ($\Delta \eta$). In the PBM this flux tube is approximated 
by a sum of partons which are distributed over this same large $\Delta \eta$
region. A blast wave gives the tubes near the surface transverse flow in the 
same way it gave flow to the bubbles in the PBM. This means the transverse
momentum ($p_t$) distribution of the flux tube is directly translated to the
$p_t$ spectrum of the PBM. 

Initially the transverse space is filled with flux tubes of large longitudinal 
extent but small transverse size $\sim$$Q^{-1}_s$. The flux tubes that 
are near the surface of the fireball get the largest radial flow 
and are emitted from the surface. As in the parton bubble model 
these partons shower and the higher $p_t$ particles escape the 
surface and do not interact. $Q_s$ is around 1 GeV/c thus the 
size of the flux tube is about 1/4f initially. The flux tubes near 
the surface are initially at a radius $\sim$5 fm. The $\phi$ angle 
wedge of the flux tube $\sim$ 1/20 radians or $\sim$$3^\circ$.
In Sec. I D (also see Sec. I C) we connect the GFTM to the PBM (see Sec. I B):
by assuming the bubbles are the final state of a flux tube at kinetic
freezeout, and discussing evidence for this connection which is further
developed in Sec. II and Sec. III.

With the connection of the PBM to the GFTM two new predictions become possible 
for our PBM. The first is related to the fact that the blast wave radial flow 
given to the flux tube depends on where the tube is initially in the transverse
plane of the colliding Au + Au system. The tube gets the same
radial boost all along its longitudinal length. This means that there is
correlated $p_t$ among the partons of the bubble. In this paper we consider
predictions we can make in regard to interesting topics and comparisons with 
relevant data which exist by utilizing the parton bubble model (PBM\cite{PBM}),
and its features related to the glasma flux tube model (GFTM)\cite{Dumitru}).

Topic 1: The ridge is treated in Sec. II.

Topic 2: Strong CP violation (Chern-Simons topological charge) is
treated in Sec. III.

We show in Sec. II that if we trigger on particles with 3 to 4 GeV/c
and correlate this trigger particle with an other charged particle of greater 
than 1.1 GeV/c, the PBM can produce a phenomenon very similar to the 
ridge\cite{Dumitru,Armesto,Romatschke,Shuryak,Nara,Pantuev,Mizukawa,Wong,Hwa}.
(See Figs. 2-8). We then selected charged particles inside 
the ridge and predicted the correlation that one should observe when compared 
to the average charged particles of the central Au + Au collisions at 
$\sqrt{s_{NN}}$ = 200 GeV\cite{centralproduction,centralitydependence}.

In Sec. II D (Comparison to data): Triggered experimental angular correlations
showing the ridge were presented at Quark Matter 2006\cite{Putschke}. Figure 7
shows the experimental $\Delta \phi$ vs. $\Delta \eta$ CI correlation for
0-10\% central Au + Au collisions at $\sqrt{s_{NN}}$ = 200; requiring one
trigger particle $p_t$ between 3 to 4 GeV/c and an associated particle $p_t$ 
above 2.0 GeV/c. The yield is corrected for the finite $\Delta \eta$ pair 
acceptance.

For the PBM generator, we then form a two charged particle correlation 
between one charged particle with a $p_t$ between 3.0 to 4.0 GeV/c and another 
charged particle whose $p_t$ is geater than 2.0 GeV/c. These are the same 
trigger conditions as in  Ref.\cite{Putschke} which is shown in Fig. 7, that
shows the corrected pair yield in the central data. Fig. 8 shows the 
correlation function generated by the PBM which does not depend on the number 
of events analyzed. The two figures were shown to be in reasonable agreement
when compared as explained previously in Sec. II D. In Fig. 9 we show the ridge
signal predicted by the PBM for very similar data but with 0-5\% centrality. 
Figure 10 shows the extraction of the jet signal. Explanations are given in 
the text.

The second prediction is a development of a predictive pionic measure of 
the strong CP Violation. The GFTM flux tubes are made up of longitudinal color 
electric and magnetic fields which generate topological Chern-Simons 
charge\cite{Simons} through the $F \widetilde F$ term that becomes a source of 
CP violation. The color electric field which points along the flux tube axis 
causes an up quark to be accelerated in one direction along the beam axis, 
while the anti-up quark is accelerated in the other direction. So when a pair 
of quarks and anti-quarks are formed they separate along the beam axis leading 
to a separated $\pi^+$ and $\pi^-$ pair along this axis. The color 
magnetic field which also points along the flux tube axis (which is parallel 
to the beam axis) causes an up quark to rotate around the flux tube axis 
in one direction, while the anti-up quark rotates in the other direction. 
So when a pair of quarks and anti-quarks are formed they will pickup or lose 
transverse momentum. These changes in $p_t$ will be transmitted to the 
$\pi^+$ and $\pi^-$ pairs.

The above pionic measures of strong CP violation are used to form correlation
functions based on four particles composed of two pairs which are opposite 
sign charge-particle-pairs that are paired and binned. These four particle 
correlations accumulate from bubble to bubble by particles that are
pushed or pulled (by the color electric field) and rotated (by the color 
magnetic field) in a right or left handed direction. The longitudinal 
color electric field predicts aligned pairs in a pseudorapidity or 
$\Delta \eta$ measure. The longitudinal color magnetic field predicts 
anti-aligned pairs in a transverse momentum or $\Delta P_t$ measure. 
The observations of these correlations would be a strong confirmation 
of this theory. The much larger unlike-sign pairs than like-sign pairs in
the PBM and the data; and the strong dip of the CI correlation at small 
$\Delta \eta$ (see Fig. 1 and Sec. III C for full details) shows very strong
evidence supporting the color electric alignment prediction
in the $\sqrt{s_{NN}}$ = 200 GeV central Au + Au collision data analyses at
RHIC\cite{PBM,centralproduction}. This highly significant dip 
($\sim$$14\sigma$) means that like-sign pairs are removed as one approaches 
the region $\Delta \eta$ = $\Delta \phi$ = 0.0. Thus this is very 
strong evidence for the predicted effect of the color electric field. The color
 magnetic anti-aligned pairs in the transverse momentum prediction, treated 
in Sec. III C as of now has not been observed or looked for. However our 
predicted specific four charged particle correlations can be used to search
for experimental evidence for the color magnetic fields.

Our success in demonstrating strong experimental evidence for the expected 
color electric field effects from previously published data suggests that 
the unique detailed correlations we have presented for searching for evidence 
for the predicted color magnetic field effects should be urgently investigated.
If we are lucky and the predicted color magnetic effects can be confirmed
experimentally we would have strong evidence for the following:

1) CP is violated in the strong interaction in isolated local space time 
regions where topological charge\cite{Simons} is generated.

2) The glasma flux tube model (GFTM) which was evolved from the color glass 
condensate (CGC) would be found to be consistent with a very significant
experimental check.

3) The parton bubble model event generator (PBM) is clearly closely connected
to the GFTM. The bubble substructure strongly supported by the PBM is likely 
due to the final state of the flux tube at kinetic freezeout.  

\section{Acknowledgments}

This research was supported by the U.S. Department of Energy under Contract No.
DE-AC02-98CH10886 and the City College of New York Physics Department. The 
authors thank William Love for valuable discussion and assistance in production
of figures.

\end{document}